\documentclass[%
 reprint,
superscriptaddress,
showpacs,preprintnumbers,
nofootinbib,
 amsmath,amssymb,
 aps,
prstper,
]{revtex4-1}

\usepackage{quoting} 
\quotingsetup{vskip=5pt} 
\usepackage[english]{babel}
\usepackage{longtable}

\usepackage{graphicx}
\usepackage{dcolumn}
\usepackage{bm}
\usepackage{hyperref}

\begin{document}


\title{Assessing the interactivity and prescriptiveness of faculty professional development workshops: The Real-Time Professional Development Observation Tool (R-PDOT)}

\author{Alice Olmstead}
\affiliation{Department of Astronomy, University of Maryland, College Park, MD 20742}
\email{aolmstead@astro.umd.edu}
\author{Chandra Turpen}
\affiliation{Department of Physics, University of Maryland, College Park, MD 20742}
\email{chandra.turpen@colorado.edu}

\date{\today}

\begin{abstract}
Professional development workshops are one of the primary mechanisms used to help faculty improve their teaching, and draw in many STEM instructors every year. Although workshops serve a critical role in changing instructional practices within our community, we rarely assess workshops through careful consideration of how they engage faculty. Initial evidence suggests that workshop leaders often overlook central tenets of education research that are well-established in classroom contexts, such as the role of interactivity in enabling student learning \cite{Freeman2014}.  As such, there is a need to develop more robust, evidence-based models of how best to support faculty learning in professional development contexts, and to actively support workshop leaders in relating their design decisions to familiar ideas from other educational contexts. In response to these needs, we have developed an observation tool, the Real-Time Professional Development Observation Tool (R-PDOT), to document the form and focus of faculty's engagement during workshops. In this paper, we describe the motivation and methodological considerations behind the development of the R-PDOT, justify our decisions to highlight particular aspects of workshop sessions, and demonstrate how the R-PDOT can be used to analyze three sessions from the Physics and Astronomy New Faculty Workshop. We also justify the accessibility and potential utility of the R-PDOT output as a reflective tool using preliminary data from interviews with workshop leaders, and consider the roles the R-PDOT could play in supporting future research on faculty professional development. 
\end{abstract}

\maketitle

\section{\label{sec:intro}Introduction}

There is a general consensus among national policy makers and education researchers that undergraduate STEM instruction can be improved through closer alignment between faculty's teaching and education research principles and findings \cite{Singer2012,PCAST2012}. For many faculty, the easiest and most efficient path towards this alignment is to adopt or adapt existing research-based instructional strategies (RBIS) in their classrooms. Consistent with this idea, many faculty professional development (PD) efforts---particularly efforts led by the discipline-based education research community---have focused on disseminating RBIS \cite{Henderson2011} and have been successful to an extent, yet we rarely critically examine these efforts to understand what contributes to these successes and how we could improve. For example, prior research has shown that physics faculty are more likely to be aware of and experiment with RBIS in their classrooms after attending the Physics and Astronomy New Faculty Workshop, but that many faculty find it difficult to persist in using these strategies over time \cite{Henderson2008,Henderson2012}. This lack of persistence could indicate both a need for long-term support and ways in which existing short-term efforts do not address faculty's needs and concerns. Scientific society leaders in biology, chemistry, mathematics, geoscience, and engineering all sponsor well-attended, discipline-specific faculty workshops akin to the New Faculty Workshop \cite{STEMWorkshopsReport2012}, and encounter similar limitations and successes \cite{Stains2015,Ebert-May2011,Borrego2010,MacDonald2005}. 

No in situ research has been conducted to directly investigate what occurs during faculty PD workshops like these large national workshops, or---to our knowledge---local, small, and/or non-discipline-specific workshops. Because of this, PD leaders have neither concrete evidence of what specific PD experiences could lead to specific outcomes \cite{Sandoval2013}, nor tools to help them justify and communicate about their design decisions in robust and consistent ways. Moreover, reports and informal evidence suggest that workshop leaders often rely on lecture to disseminate ideas about teaching with minimal participant contributions \cite{Mulnix12}, which leads us to think that workshops could be improved through careful consideration of how alternative, more engaging design choices might produce more desirable shifts in faculty's thinking, practice, or participation in ongoing PD \cite{Derting2016,Loucks-Horsley2009,Desimone2002,Garet2001a,Wilson2013a,Valli2002}. 

Despite the lack of in situ observation and analysis of workshops, prior research can lend some insight into what could be contributing to the limitations of workshops (beyond time constraints), and what could be done to address this. Looking across 191 published approaches to faculty PD in STEM, \textcite{Henderson2011} find that faculty change efforts typically fall into one of four categories: disseminating curriculum and pedagogy, enacting policy, developing reflective teachers, and developing shared vision. These categories are distinguished from each other by whether they target prescriptive or emergent outcomes, and whether they target individuals or groups. Workshops naturally fall into the prescriptive, individually-focused ``dissemination'' category, and often encourage faculty to adopt strategies as-is. However, many faculty often want or need to modify RBIS to fit their local contexts, which is not well-addressed by a one-way communication model. Compounding this problem, highly prescriptive professional development may make it difficult for faculty to feel that their experiences and insights matter, which could limit their future engagement with education research communities \cite{HendersonDancy2008}. These shortcomings of the disseminating curriculum and pedagogy approach are well-addressed by the change strategies of developing shared vision or reflective practice. These approaches to PD, which are more focused on emergent outcomes, are likely critical to faculty's thoughtful adaptation and sustained use of RBIS. Although the four change strategies defined by \textcite{Henderson2011} tend to be fairly siloed, we think that workshops could support emergent outcomes by engaging faculty in the kinds of reflection around instruction we would want them to engage in when they teach, and by explicitly helping faculty to identify ways to engage with communities of educators at their home institutions.

In order to support the exploration and analysis of various approaches to PD workshop design and implementation, we have developed an observation tool, the Real-time Professional Development Observation Tool (R-PDOT), which can broadly document what faculty experience when they attend PD workshops. In this paper, we discuss the development of and intentions behind the R-PDOT and demonstrate how its output can be used to reflect on workshop design. The structure of our paper is as follows. In Section \ref{sec:theoreticalapproach} we describe the theoretical framework that underlies this work.  In Section \ref{sec:background}, we overview research on reflective teaching discussions that informs the structure of and PD elements captured by the R-PDOT. In Section \ref{sec:tooloverview}, we define and justify the specific descriptive codes that comprise the R-PDOT. In Section \ref{sec:methods}, we consider the accessibility of the R-PDOT, including our methods for refining and using the R-PDOT codes, establishing inter-rater reliability, and visualizing data. In Section \ref{sec:analysis}, we analyze R-PDOT data from three sessions of the Physics and Astronomy New Faculty Workshop in order to demonstrate how the R-PDOT data can provide a basis for hypothesizing what kinds of faculty outcomes these different designs might support. In Section \ref{sec:discussion}, we elaborate on the potential implications of trends we noticed by using the R-PDOT in the Physics and Astronomy New Faculty Workshop, and show preliminary evidence that the R-PDOT data can support workshop leaders in noticing similar key features during discussions about workshop design. We conclude in Section \ref{sec:conclusions} by considering how this research could enable members of the education research community to engage in critical conversations about workshop design, and how the R-PDOT could enable future researchers to generate and pursue new research questions about PD. 

\section{\label{sec:theoreticalapproach}Theoretical approach}
Broadly speaking, we take a Vygotskian stance that different ways of thinking, seeing, and knowing would likely appear first in distributed form, i.e., in interactions with others, and then internalized by individuals (and transformed in the process). These interactions can provide the opportunity for a diversity of ideas to be contested, compared, and developed in ways that build meaning and allow individuals to begin to take up perspectives that may have been different from their initial ideas. This theoretical approach is broadly applicable to all learners, be it students, K-12 teachers, faculty, or workshop leaders. 

It would be difficult to meet many learners---with a diversity of incoming knowledge and experiences---where they are at through a one-size-fits-all communication approach, like lecture. It is well-recognized within the education research community that active engagement is a positive attribute of classroom teaching, as demonstrated by the vast number of studies within undergraduate STEM education linking increased active engagement to outcomes such as higher student conceptual gains and decreased failure rates \cite{Freeman2014}. The positive impacts of active learning have also been demonstrated within teacher PD, where increased active engagement within PD is strongly correlated with greater improvements in teacher practice \cite{Garet2001a}. Though similar empirical evidence does not yet exist in higher education PD, we assume the benefits of active engagement extend to faculty learning.

We recognize that the construct  \emph{active engagement}  is too vague to fully inform the delineation of different types of engagement: both researchers and instructors use this term to refer to a wide variety student interactions and behaviors \cite{Dancy2007}, and some of these are may be more or less valuable for learners. From a Vygotskian perspective, the forms of faculty interactions are consequential to what faculty would learn from these experiences. In considering the form of faculty's interactions, we instead rely on research surrounding \emph{participant structures}, namely, the ``configurations of interactional rights and responsibilities that arise within particular activities'' \cite{OConnor1993,Goodwin1990}. Whereas researchers often do not distinguish between variations in types of active engagement, well-defined variations within participant structures have been extensively explored within the literature  \cite{Jurow2005,OConnor1993,Knuth2001,Lotman1988,Bakhtin1986,Scott2006}. We revisit this literature when describing the specific form of the R-PDOT in Section \ref{sec:tooloverview}.

\section{\label{sec:background}Background Literature}
We intend the R-PDOT to capture a range of prescriptive to emergent intended faculty PD outcomes. In creating the tool, we recognize that a variety of PD goals exist, from straightforward goals of raising faculty's awareness of what resources and materials exist, to more ambitious goals of improving faculty's abilities to notice student thinking, reflect on their instruction, and engage in future collaborative discussions around instruction. We relate these more ambitious goals to emergent outcomes, in that the final form of faculty's instruction is not predetermined. We consider awareness-level goals to be easy for workshop leaders to achieve through a variety of mechanisms, including lecture, and therefore do not explore this approach in depth in this background section. In contrast, for more ambitious goals that target emergent outcomes, different kinds of faculty engagement are likely needed, and the form of faculty's interactions matter for what they take away from workshop experiences. We use this background section to explore some PD characteristics that could support ambitious outcomes.

Faculty PD studies are arguably the most relevant to understanding faculty's learning about teaching. Because the participant characteristics in these studies mostly closely match the characteristics of our target population, the claims we make that draw directly from this literature are likely the most robust. Faculty's teaching is often constrained by a variety of factors, including departmental organizational structures, college accreditation, and competing expectations of research excellence, which are not at play in K-12 settings. However, there are goals of faculty PD that significantly overlap with K-12 teacher PD goals, and the endeavor of teaching is fundamentally similar in many ways across undergraduate and K-12 teaching.  Both faculty and K-12 teachers likely have routines of practice informed by their prior teaching experiences, and therefore would likely face some common struggles in shifting their thinking and practice, such as learning how to listen to and foster students' potentially productive disciplinary ideas instead of expecting to ``fix'' or replace students' wrong ideas through lecture. In addition, K-12 teacher PD researchers have made strides in areas that faculty PD researchers have not: for instance, in constructing claims about the nature of instructors' conversations about teaching and what PD activities are often linked to shifts in instruction. While future research that examines faculty's interactions directly would strengthen our claims (since differences between faculty and K-12 teacher's incoming knowledge and preparation may necessitate additional or different support structures to produce similarly productive interactions), currently, the synthesis of this range of literature best allows us to elaborate on how and why certain PD outcomes could reasonably be linked to specific workshop activities.

This literature also informs how we think about supporting workshop leaders in improving their PD practice, which is the overarching goal of creating this workshop observation tool. Analogous to how PD aims to support teachers in fostering student learning in new ways, we aim to support workshop leaders in fostering faculty learning in new ways. Because there is no prior research about how workshop leaders develop their ideas about how to design and implement PD, our tool design decisions are often directly influenced by the literature referenced above (without additional comparisons to literature about workshop leaders' learning or interactions). We suggest that there are sufficiently strong similarities between workshop leaders, faculty, and K-12 teachers, both in the ways they could develop new ways of thinking and noticing through structured discussions with others, and in the ways they could act as facilitators in formal, classroom-like settings, to justify drawing from faculty and K-12 teacher PD research when we consider how to support workshop leaders' thinking about workshop design. 

Because we consider reflective discussion around examples of instruction to be a central mechanism for generating new ways of thinking and knowing for teachers (in this case, faculty and workshop leaders), we begin by reviewing relevant faculty and K-12 teacher PD literature on this topic in the following section.

\subsection{Overview: Cultivating Reflective Discussion}
Here, we review faculty and K-12 teacher PD literature that illuminates the potential impacts of reflective discussions about instruction and the supports that seem necessary for these conversations to be productive. As mentioned above, the literature on reflective teacher discussions influences the development of our observation tool in two important ways: (1) it informs specific aspects of effective faculty PD that we choose to capture with the R-PDOT (described in more detail in \ref{sec:tooloverview}); and (2) it informs the overall form of the R-PDOT, as we intend the tool to serve as an effective vehicle for reflective discussion and debate among workshop leaders.

In both of these cases, reflective discussion can be a critical part of improving instructional practice and preparing instructors for future learning. Yet it would be na\"{\i}ve to encourage any discussions about instruction and expect consistently desirable outcomes. Not all reflection is equally valuable; instead, the form of teacher reflection matters for the degree of resultant instructional improvement. It is all too common for teacher reflection to be insufficiently structured and poorly supported \cite{HammersleyFletcher2005,Chism2007}. For example, faculty peer observations of teaching often focus on superficial tips and tricks that are unlikely to improve student outcomes in a significant way even if taken up by faculty \cite{Hammersley-Fletcher2004}. These observations often do not challenge faculty's initial notions of what constitutes effective peer review, and lack concrete examples or other information that could help faculty to operationalize overly general guidelines \cite{Chism2007}. Similarly, without guidance, pre-service K-12 teachers often struggle to produce written reflections that adequately prepare them to change aspects of their practice, focusing only on what occurred in the classroom but not why it may have occurred or what lessons they might take away from that experience \cite{Aubusson2010}.  All of these minimally useful observations tend to be highly unstructured, where we use the term \emph{structures} to denote both verbal and written scaffolding that can direct teachers to assess certain aspects of instruction or student engagement. 
%
%
%
%
%
%

In the other extreme, if support structures are perceived as too evaluative or judgmental, the likelihood of instructional change can decrease. Specifically, rubric-based observation tools (though helpful in some ways) can create significant barriers to promoting thoughtful reflection and encouraging self-motivated improvement due to their evaluative nature. Teacher resistance to this evaluation is justifiable. At the K-12 level, rubrics are often used for formal teacher assessment rather than for reflective self-improvement, and the outcomes contribute to decisions about career advancement or setbacks. Observers' assessments may be less consistent in practice than anticipated by the designers, different rubrics may yield different assessments of teaching quality for the same class, and there is often significant variation in scores from class-to-class for individual teachers, yet high-stakes assessments are sometimes based on only one observation \cite{Kane2011,Guarino2012,Amrein-Beardsley2011}. Even at the college level, where the stakes are typically quite low, the highly critical feedback that is associated with the implementation of these rubrics often shuts down conversations with instructors and makes them less willing to engage in PD efforts instead of sparking productive conversations around teaching \cite{HoraFerrare2013,Chism2007}. 

\subsection{Structures for cultivating reflective discussions}
In order to identify some key features that can support both critical and transformative reflections on teaching, here we consider prior research that shows evidence of effective support structures. Many prior studies have demonstrated that teachers can improve when guided to reflect in ways that help them to identify meaningful aspects of their practice to change or maintain \cite{Amrein-Beardsley2011,Gallos2005,Morrell2012,Aubusson2010,MacIsaac2001,Hampton2004,McShannon2006,Hativa1995,Piccinin2006}, and these reflective PD efforts have common elements that motivate the design of our observation tool. 

In particular, we find that \textcite{Aubusson2010} provide useful terminology that allows us to articulate what kinds of scaffolding can encourage reflection that leads to productive changes to instruction. The authors investigate how pre-service teachers learn to become reflective and what supports them in doing so, and conclude that a focus on \emph{contextual anchors}---specific teacher practices that are  observed first-hand (as the instructor) or by watching others---was critical for making the pre-service teacher's reflections generative for their future instruction. They also posit that a focus on \emph{conceptual anchors}---direct connections to education research theory---may be equally important in strengthening teacher reflection. In other words, \textcite{Aubusson2010} show that when teachers are able to leverage both observed student behaviors and prior education research when reflecting on instruction, they are able to take next steps that improve their teaching and their students' outcomes. 

Other studies in K-12 teacher PD corroborate \textcite{Aubusson2010}'s claims. \textcite{HornLittle2010} find that teacher conversations that routinely focus on specific problems encountered in teaching and connect these specific examples to general principles of teaching and learning are more likely to generate viable solutions than conversations that do not. \textcite{VanEs2002,VanEs2008} take a similar stance by defining the ability to connect teaching events to general principles as one of the three key features of teacher \emph{noticing}, and subsequently show that teachers improve their practice as they become more adept at noticing \cite{Sherin2008,VanEs2010}. In addition to the similarities in what individual reflective practices were developed across these studies, we also note here that all of these studies centrally involve discussion with other instructors, and consider this to be an important piece of how teachers' abilities develop over time.

We can use these ideas to see connections between other studies that involve similar PD elements but do not provide such detailed, mechanistic explanations for their successes. For one, many reflective PD efforts are facilitated by knowledgeable coaches who offer feedback to instructors, and they attribute much of the positive shifts in faculty's attitudes, instruction, and/or student outcomes to this facilitation \cite{Hampton2004,Hativa1995,Piccinin2006,McShannon2006,Gallos2005}. Although PD researchers are often not explicit about what is discussed during instructor-facilitator consultations, it is highly likely that the facilitators, who are sometimes the researchers themselves, are making connections between instructors' current practices and education research theory, and using these connections to centrally inform how they guide instructors to reflect. Moreover, an argument for the importance of finding contextual anchors is consistent with an argument for the importance of teachers and PD facilitators gaining shared knowledge of classroom events, which occurs in many reflective PD efforts \cite{Piccinin2006,Gallos2005,McShannon2006,Hativa1995}. In other words, when facilitators learn about how instruction plays out by conducting classroom observations, they are better able to help faculty identify both contextual and conceptual anchors, which together can be used to inform highly productive changes to faculty's instruction.

Structured observation tools or protocols can further enable productive discussions around instruction by providing a guide for the observer as they decide what written and verbal feedback to supply, and a way for the teacher being observed to independently recall key aspects of what occurred in their class based on what is written down, rated, and/or tallied. Rubric-based observation protocols set evaluation criteria that are supported by education research theory, and therefore can encourage both highly experienced and new educators to make connections between classroom events and theoretical ideas \cite{Amrein-Beardsley2011,MacIsaac2001,Morrell2012}. For example, in \textcite{MacIsaac2001}'s study, pre-service and inservice teachers used the Reformed-Teaching Observation Protocol (RTOP) \cite{Sawada2002} to analyze teaching segments, and became better able to articulate what about their teaching should shift and why. The authors argue that much of this improvement can be attributed to the teachers becoming familiar with the RTOP items and better able to relate them to how science students were engaged during real events. Similarly, \textcite{Morrell2012} explicitly take up \textcite{Aubusson2010}'s framework and show evidence that their rubric-based observation protocol, the Oregon Teacher Observation Protocol (OTOP) \cite{Wainwright2003}, enabled pre-service teachers to identify conceptual and contextual anchors and led to focused, productive reflection on instruction.

As mentioned at the start of Section \ref{sec:background}, parallels exist between workshop and classroom contexts that lead us to consider how workshop leaders make decisions as educators. Just as classroom educators must balance many constraints and define goals as they decide how to guide students to engage with disciplinary practices and ideas, workshop leaders must navigate their constraints and choose among potential goals as they design workshop sessions for faculty. Thus, the same attributes that characterize productive reflection on classroom instruction should inform how we support workshop leaders' reflection on workshop sessions. When it comes to improving workshop leaders' abilities to reflect, we think that a descriptive (non-rubric-based) observation tool has the greatest potential to promote fruitful, reflective discussions that are anchored to both education research theory and workshop session events without creating tension between workshop leaders and observers. 

With descriptive/non-evaluative classroom observation tools, an observer captures the minute-by-minute flow of events through a set of codes that describe teacher and/or student behaviors, but does not assign a specific score to the class based on the prevalence of these codes \cite{Smith2013,Hora2013,West2013}. This is the approach we have chosen for the R-PDOT. As the creators of the tool, we highlight aspects of workshop sessions that we consider to be strongly connected to key workshop outcomes, and in doing so implicitly create connections to education theory and our vision for professional development \cite{HoraFerrare2013,Goodwin1994}. In particular, the R-PDOT data helps to portray workshop events in the way that we see them, similar to the processes of \emph{highlighting} and \emph{coding} that \textcite{Goodwin1994} defines as essential, often subconscious mechanisms by which people develop shared vision. In an educational context, \textcite{VanEs2002,VanEs2008} tie \textcite{Goodwin1994}'s general depiction of developing professional vision to the ways that teachers learn to notice significant teaching events and connect them to theory, and we imagine the R-PDOT data functioning similarly here. The R-PDOT codes also employ language that is likely to bring up foundational aspects of teacher PD that could facilitate workshop leaders gaining new insights into their PD practice, as has been argued of providing rich language for teachers to reflect on their instruction \cite{Scott2006,MacIsaac2001,MacIsaac2002,Wainwright2004}. 

Despite this implicit guidance, the descriptive nature of the tool gives workshop leaders agency to interpret the data in multiple ways and determine their own next steps, which has been shown to be important for catalyzing productive discussions about instruction \cite{HornLittle2010}. The nature of the R-PDOT data allows simple visual representations of results, which can further enable workshop leaders to make sense of the data for themselves, as we discuss in Section \ref{sec:visualization}. We note that we are also well-justified in expecting the R-PDOT to serve as a research tool in future studies based on the long history of classroom observation tool use in the context of teacher education research and assessment \cite{HoraFerrare2013,Kane2011,Turpen2009,Turpen2010,Adamson2003,Amrein-Beardsley2011,Budd2013,Smith2014,Morrell2004,Walkington2013,Stang2014,Lund2015,Wainwright2004}. 

Many of the aspects of workshop sessions we choose to foreground with the codes themselves link back to this same literature about reflective teaching: the codes allow a user to differentiate times when faculty could be engaged in activities that mirror the productive discussions described above from times when faculty are not deeply immersed in pedagogical discussions. Specifically, R-PDOT codes can capture when faculty are analyzing concrete examples of instruction, when connections between instruction and education research might be articulated, and when multiple faculty perspectives might be voiced, debated, and discussed. We define and justify these codes in the following section.

\section{\label{sec:tooloverview}Tool Overview}
Here, we define the descriptive codes that comprise the R-PDOT and elaborate on why each code was created. The R-PDOT codes broadly capture the form and focus of faculty's engagement during PD workshops, and encompass most common PD approaches (as discussed in more detail in Section \ref{sec:methods}). Generally speaking, the R-PDOT has a similar form and function to classroom observation tools such as the Classroom Observation Protocol for Undergraduate STEM (COPUS) \cite{Smith2013}, the Real-time Instructor Observation Tool (RIOT) \cite{West2013}, and the Teaching Dimensions Observation Protocol (TDOP) \cite{Hora2013}: it allows an observer to collect non-evaluative, quantitative data about what faculty participants experience during workshops. An observer can simultaneously capture two complementary aspects of workshop sessions---the ways in which faculty participants are engaged and the focus of their engagement---with two sets of codes, and thus highlight aspects of workshop sessions that prior research suggests may lead to particular kinds of outcomes. Each set of codes is intended to fully span the ways that faculty are most likely to be engaged or have their attention focused during workshop sessions.

For practical reasons, the R-PDOT codes are fairly broad in scope, in the sense that a single code could be enacted in a variety of ways.  By defining the R-PDOT codes, we necessarily foreground large-scale characteristics of workshop sessions that an observer can quickly record with reasonable fidelity (as justified in Section \ref{sec:methods}). In doing so, we lay the groundwork for reflection and follow-up analysis that targets more nuanced variations. In general, the prevalence or absence of a particular session focus, and the type of faculty engagement paired with that focus, should inform workshop leaders about what faculty outcomes are more or less plausible from a given session. For example, it would be implausible to expect a workshop session in which a workshop leader primarily lectures about education research results to improve faculty's abilities to write conceptual questions for their students; whereas a session in which faculty primarily practice writing such questions might. However, in this second example, the nature of the question-writing task, the quality of any written scaffolding the workshop leader provides, and the exact facilitation moves the workshop leader employs in the moment will also influence workshop outcomes, and these are aspects of PD environments that the R-PDOT is not designed to characterize. 

As we describe the codes in this section, we elaborate on a few significant potential variations within individual codes in order to encourage readers to think critically about how each one is enacted. We also note that although we typically consider each code separately in this section, we also encourage workshop leaders to consider how codes are combined when assessing real workshop session design, as we illustrate in Section \ref{sec:analysis}. 

\subsection{\label{sec:typeofengagement}Type-of-Engagement Codes}
The first dimension the R-PDOT allows an observer to capture is the type of faculty's engagement during a workshop session, i.e., whether faculty are listening to lecture, working independently, engaged in small group work, or engaged in some sort of large group discussion at any given time. By drawing attention to the type of engagement within faculty PD workshops and thus enabling discussions around active engagement in workshops, we clearly align our tool with the accumulated knowledge and interests of the discipline-based education research community. These codes are defined in Table \ref{tab:typeofengagement}, and an extended codebook (with examples) can be found in Appendix \ref{appendixA}.

\bgroup
\def\arraystretch{1.5}
\begin{table*}
\caption{\label{tab:typeofengagement} Type-of-engagement code names and brief descriptions.}
\begin{ruledtabular}
\begin{tabular}{p{0.35\textwidth}p{0.6\textwidth}}
\emph{Code name} & \emph{Code description}\\
\hline
Workshop Leader Lecture & Workshop leader lectures while faculty participants listen.\\
Large Group Closed,\newline Faculty Participant Question & Large group closed discussion. Faculty participant(s) question the workshop leader, and (optionally) the workshop leader answers through lecturing, while all other participants listen.\\
Large Group Closed,\newline Workshop Leader Question & Large group closed discussion. Workshop leader asks non-rhetorical closed questions, and (optionally) one or more faculty participants respond directly to the workshop leader\cite{Dancy2007}.\\
Large Group Open Discuss & Large group open discussion. Workshop leader and faculty participants take turns speaking, with the discourse focused on the ideas of faculty participants\cite{Dancy2007}.\\
Small Group Discuss & Faculty participants discuss with each other in small groups.\\
Faculty Participant Present & One or more faculty participants present to all others.\\
Faculty Participant Independent Work & Faculty participants work independently on a task.\\
\end{tabular}
\end{ruledtabular}
\end{table*}
\egroup

Many classroom observation tools capture aspects of participant structures that we also target \cite{West2013,Marshall2010,Lane2015,Gutierrez1999,Hora2013}. One common delineation in the participant structure-oriented literature is the distinction between \emph{closed} and \emph{open discussion} \cite{OConnor1993,VanZee1997,Scott2006,Turpen2009}, which has been used explicitly by other observation tool developers \cite{West2013,Gutierrez1999}. \emph{Closed discussion} refers to instances where a teacher guides students towards a single, predetermined correct answer, frequently following a pattern of I-R-E: teacher initiation, student response, and teacher evaluation \cite{Cazden2001,Mehan1979,Lemke1989}. While closed discussion has the advantage of ensuring that canonical knowledge is voiced within the classroom, it can be problematic when used extensively because it positions the teacher as the sole authority and suppresses students' agency to explore, develop, and defend their own ideas \cite{VanZee1997,Lemke1989,Dancy2007}. In contrast, \emph{open discussion} refers to instances where a teacher encourages student contributions while withholding their own evaluation or judgment and often asks questions that do not have predetermined answers, which supports students' development in ways that closed discussion does not. 

Researchers who have studied classroom discourse have argued that a well-balanced and well-ordered combination of different participant structures can improve learners' abilities to engage in disciplinary conversations and other practices outside of formal settings \cite{Scott2006,Engle2010,Schwartz1998}. We share a commitment to the importance of such outcomes when considering the engagement of any learners, undergraduate students and faculty alike, which further justifies our attention to these constructs in creating our type-of-engagement codes. When it comes to teacher PD specifically, prior research suggests that it can be productive when more experienced or pedagogically knowledgeable teachers position others as having agency in changing their own instruction instead of prescribing exact solutions \cite{HornLittle2010}, which underscores the potential value of open discussion in workshop contexts.

\bgroup
\def\arraystretch{1.5}
\begin{table*}
\caption{\label{tab:communicativeapproaches} This table locates the R-PDOT type-of-engagement codes within the four classes of communicative approaches---non-interactive/authoritative, non-interactive/dialogic, interactive/authoritative, and interactive/dialogic---defined in \citet{Scott2006}, Table 3, p.611. These distinctions are discussed further in the main text.}
\begin{ruledtabular}
\begin{tabular}{p{0.25\textwidth}|p{0.4\textwidth}|p{0.35\textwidth}}
& \emph{Authoritative}: purpose is to focus participants on one meaning & \emph{Dialogic:} open to many points of view\\ \hline
\emph{Non-interactive:} excludes contributions of other people & Workshop Leader Lecture & Faculty Participant Independent Work\\ \hline
\emph{Interactive:} allows contributions of more than one person & Large Group Closed, Workshop Leader Question\newline Large Group Closed, Faculty Participant Question & Faculty Participant Present\newline Small Group Discuss\newline Large Group Open Discuss\\
\end{tabular}
\end{ruledtabular}
\end{table*}
\egroup

\textcite{Scott2006}'s analytical framework for distinguishing between participant structures closely matches our approach and allows us to articulate key differences between our type-of-engagement codes, as shown in Table \ref{tab:communicativeapproaches}. The authors define four classes of communicative approach using two dimensions: \emph{authoritative} versus \emph{dialogic}, and \emph{non-interactive} versus \emph{interactive}. \emph{Authoritative} communicative approaches aim to focus participants on one meaning while \emph{dialogic} communicative approaches are open to many points of view; \emph{non-interactive} communicative approaches exclude contributions of other people while \emph{interactive} communicative approaches allow the contributions of multiple people.

Within our codes, non-interactive/authoritative is exemplified by ``Workshop Leader Lecture'': a workshop leader presents a single point of view and faculty participants are excluded from contributing. We also agree with \textcite{Scott2006} that lecture can be dialogic in principle, meaning that a lecturer could present, compare, and contrast participants' ideas without evaluating them, but we have found this mode of lecture to be rare within workshop settings and therefore assume lecture is authoritative unless an observer notes otherwise. In our table, we associate ``Faculty Participant Independent Work'' with the non-interactive/dialogic communicative approach instead. Here, each faculty participant has freedom to generate their own ideas and there is typically no opportunity for others to contribute to the development of those ideas. 

Within the interactive/authoritative communicative approach, ``Large Group Closed, Workshop Leader Question'' maps clearly onto \textcite{Scott2006}'s definition, which encompasses both I-R-E-type questioning and longer chains of questions where it is evident that the workshop leader is looking to develop one particular idea and not others. By our definition, ``Large Group Closed, Faculty Participant Question'' differs from ``Large Group Closed, Workshop Leader Question'' in that a workshop leader is answering faculty participant questions instead of the reverse. Because the workshop leader is positioned in an authority role, faculty participant questions typically lead a workshop leader to respond based on their own thinking, which maintains this participant structure as we have defined it and establishes the full exchange as authoritative. We note that these periods of addressing faculty participant questions can be minimally interactive.

The dialogic/interactive communicative approach encompasses three type-of-engagement codes: ``Faculty Participant Present,'' ``Small Group Discuss'' and ``Large Group Open Discuss''. We again find \textcite{Scott2006}'s approach useful to explain what we see as meaningful differences between these codes and justify these delineations. Using \textcite{Scott2006}'s framework, ``Faculty Participant Present'' represents a \emph{low level of interanimation of ideas}: multiple perspectives are simply made available in the public space when faculty participants present to the whole group, and there is no opportunity for consensus to be reached or differences understood if faculty voice conflicting ideas. While it could be sufficient for faculty to share ideas without discussion or debate, learning opportunities might also be missed if faculty presentation is not followed up by a communicative approach that encourages comparison. In large group open discussion and small group discussion, many ideas may be compared, contrasted, contested, and developed in relation to each other---a \emph{high level of interanimation of ideas}. Following the same reasoning highlighted by \textcite{Scott2006}'s framework, ``Small Group Discuss'' could have both drawbacks and affordances relative to ``Large Group Open Discuss'':  in ``Small Group Discuss,'' only a limited number of faculty participants have access to any particular group's ideas, but there are likely increased opportunities for each participant to contribute.

We return to how our real-time application of these codes compares to \textcite{Scott2006}'s intent and methodology in Section \ref{sec:methods}.

\subsection{\label{sec:focusofengagement}Focus-of-Engagement Codes}
\subsubsection{Defining focus-of-engagement codes}
Here we define and justify the second dimension of the R-PDOT codes: the focus of faculty participants' engagement. By focus-of-engagement, we mean the topical focus of the workshop session that faculty are asked to think about or engage with, e.g., education research results, abstract descriptions of instructional strategies, concrete examples of research-based instructional strategy implementation, or the past experiences of people within the workshop. Table \ref{tab:focusofengagement} contains a complete list of the focus-of-engagement codes and brief descriptions of what they encompass. We note that unlike the type-of-engagement codes, where we expect to be able to capture the behavior of a majority of faculty participants with a single code, more than one focus-of-engagement code can occur simultaneously for faculty---these categories are not exclusive. (An extended codebook with examples of each code can be found in Appendix \ref{appendixB}, and we discuss our coding methodology further in Section \ref{sec:methods}.)

\bgroup
\def\arraystretch{1.5}
\begin{table*}
\caption{\label{tab:focusofengagement} Focus-of-engagement code names and brief descriptions.}
\begin{ruledtabular}
\begin{tabular}{p{0.35\textwidth}p{0.6\textwidth}}
\emph{Code name} & \emph{Code description}\\
\hline
Workshop Instructions & Workshop leader instructs participants about what they should or will be doing during the workshop (or participants attempt to clarify these instructions).\\
Education Research Theory and Results & Workshop leader and/or participants emphasize discipline-based education research processes, principles, or findings.\\
Instructional Strategies (IS) Description and Purpose & Workshop leader and/or participants show or describe active learning strategies ranging from current instructional practices to strongly research-based instructional strategies.\\
Workshop Leader (WL) Simulating IS & Faculty participants experience a workshop leader's implementation of an instructional strategy, either by acting as mock students, or through observing video, transcript, or case study narrative.\\
Faculty Participant (FP) Simulating IS (as educator) & A predetermined subset of faculty participants (one or more) try out implementing an instructional strategy while other participants act as mock students.\\
Analyzing Simulated IS & Workshop leader and/or participants reflect on (analyze, critique, evaluate, justify) a shared experience of someone simulating an instructional strategy in situ.\\
WL Pre-Workshop Experiences & Workshop leader and/or participants discuss a workshop leader's past experiences, including instructional goals, practices, values, and local contexts.\\
FP Pre-Workshop Experiences & Workshop leader and/or participants reflect on participants' past experiences, including instructional goals, practices, values, and local contexts.\\
Student Experiences & Workshop leader and/or participants consider students' knowledge, skills, or affect.\\
Disciplinary Content Knowledge & Workshop leader and/or participants consider disciplinary ideas.\\
Analyzing and/or Creating Student Tasks & Participants create, modify, or evaluate and/or workshop leaders critique or evaluate specific materials, questions, or tasks for students.\\
Planning for FP Future Teaching & Workshop leader advises and/or faculty participants plan next steps for when participants go back to their home institutions.\\
\end{tabular}
\end{ruledtabular}
\end{table*}
\egroup

\subsubsection{Justifying focus-of-engagement codes}
In the remainder of this section, we consider the potential outcomes associated with enacting each focus-of-engagement code. For simplicity, we do not consider in depth what type of engagement and activity duration would plausibly generate these outcomes. However, we again emphasize that the type of faculty engagement is likely to be highly consequential to what they learn, and throughout these codes, we expect the most ambitious outcomes to emerge if there are extended periods that exemplify a dialogic/interactive communicative approach, where faculty's ideas are explored and developed. Therefore we encourage PD leaders to consider the intersection of the two R-PDOT dimensions when using this tool to inform future workshop design, as opposed to considering the potential value of the workshop foci in isolation from the extent to which faculty are engaged in making sense of relevant ideas.

\emph{Workshop instructions:} A workshop leader's instructions have the potential to shape faculty participants' engagement and learning throughout the session. While faculty likely bring their own incoming expectations when they enter a workshop or session, a workshop leader can intentionally try to cue up certain ideas that might otherwise be dormant by telling faculty how to approach a particular workshop task or what they are expected to gain from attending a particular session. \citet{Hammer2005} describe the general process by which people relate their current situation to familiar past experiences as \emph{framing} or resource activation, and several studies have shown that students' engagement and apparent abilities are influenced by the framing they adopt \cite{Smith1994,diSessa1998,Scherr2009}. Similarly, PD researchers have shown that instructors approach teaching in context-dependent ways \cite{Markauskaite2014,Hora2012,Cohen2003,Stroupe2013,Harlow2013}, which lends weight to the idea that the framing that faculty take up within workshop settings plays a role in what they gain from their participation. Although faculty may quietly or vocally contest a workshop leader's instructions, every participant will adopt some framing that influences their thinking during the session, regardless of whether or not the workshop leader guides them towards a particular orientation. Because faculty participants and workshop leaders might not naturally take up the same framing, a session in which the workshop leader's desired framing is made explicit is likely to foster stronger alignment between the workshop leader's expectations and participants' actual experiences than a session where no expectations are articulated.

\emph{Education Research Theory and Results:}
Unpacking the theoretical motivations that informed the development of research-based instructional strategies (RBIS) could help faculty to decide how and when to adapt, modify, or reinvent them. Prior research has shown that physics faculty often struggle to make these decisions in ways that both support positive student outcomes and fit within their local constraints, and would likely benefit from a deeper understanding of the guiding principles that underlie the developers' prescriptions \cite{Dancy2009PERC}. More generally, faculty's ability to relate education research theory to examples of classroom practice is critical to their ability to continually assess and improve their own teaching, as discussed previously \cite{Horn2010,Aubusson2010,Morrell2012}. A focus on education research theory in close proximity to the ``Analyzing Simulated IS'' code would likely indicate that workshop leaders or faculty participants are identifying the conceptual anchors of  \textcite{Aubusson2010} that tie theory to practice, which we consider to be a particularly promising and rich PD activity. 

Working towards an alternative goal, introducing faculty participants to education research results that are new to them might improve their ability to try innovative strategies in environments where institutional or departmental pressures constrain their teaching. In particular, physics faculty could justify their decision to use active learning strategies to resistant administrators using education research findings and methods, e.g., using concept inventories to measure student outcomes \cite{Henderson2014,DancyPERC2010,Turpen2016}. That said, we caution that contrary to popular belief, quantitative research results often do not contribute to convincing individual faculty to initially try RBIS \cite{HendersonDancy2008,DancyPERC2010}, and we argue that a workshop that focuses on this exclusively is unlikely to shift faculty participants' willingness to try new teaching strategies.

\emph{Instructional Strategies Description and Purpose:}
The simplest way for faculty to become aware of RBIS is to hear them described. Prior research on the Physics and Astronomy New Faculty Workshop suggests that workshop presentations are an effective method of introducing faculty to a variety of RBIS, thus expanding faculty's awareness of what they could do in their classrooms \cite{Henderson2008,Henderson2012}. Workshops could be designed to increase participants' familiarity with popular RBIS that they seem likely to take up, particularly when time is limited. Alternatively, workshop leaders could elicit instructional strategies from participants, who may contribute a greater diversity of strategies than a workshop leader would have presented based on thoughtful planning, but lead to a higher likelihood of participants attempting RBIS in their classrooms because the strategies were endorsed by their peers \cite{DancyPERC2010}. Although describing instructional strategies is likely to be insufficient for preparing faculty participants to navigate the challenges of implementing them in their own classrooms, and indeed, prior research indicates many will not persist in using RBIS in the long-term after only this intervention \cite{Henderson2012}, it can give faculty a valuable starting point for experimentation within their classrooms. 

\emph{Workshop Leader Simulating Instructional Strategy:}
Unlike describing instructional strategies in the abstract, simulating instructional strategies can give faculty participants concrete models of instruction to reflect back on. This could serve to make the nuances of RBIS implementation visible to faculty participants and more richly illustrate the possible impacts for students than a simple description \cite{Prather2008,Turpen2009}, which may in turn increase the likelihood that some faculty will implement similar strategies when they return to their home institutions. K-12 PD designers have argued that simulating instructional strategies for participants can make PD seem more authentic and therefore make new strategies seem more plausible \cite{Loucks-Horsley2009}, and indeed, empirical evidence supports the existence of a causal link between K-12 teachers experiencing instructional strategies in workshops and later shifting their own instruction accordingly \cite{Garet2001a,Desimone2002,Darling-Hammond2009}. We note that although the three forms of simulating instructional strategies that we group together with this code---faculty participants acting as pseudo-students while workshop leaders model instructional practices; workshop leaders using video-recorded classroom episodes to engage faculty participants in classroom implementation; and faculty participants reading case studies of classroom interactions---can all provide the contextual anchors necessary for productive reflection \cite{Aubusson2010,Horn2010,VanEs2002,VanEs2008}, each may have affordances that the others do not. We encourage future researchers and PD designers to explore the consequences of these variations through follow-up analysis, but choose not to distinguish between them in this initial classification.

\emph{Faculty Participant Simulating Instructional Strategy (as educator):}
Asking faculty participants to enact instructional strategies themselves could benefit them in many similar ways to what is described in the previous code, but there are also sufficient differences in the potential workshop outcomes to justify making this a separate code. In particular, while workshop leaders' simulations can provide a limited number of ``expert-like'' models of instruction, having many faculty participants simulate teaching strategies could provide a greater diversity of concrete, shared examples of practice that can support rich reflective discussion and debate (again, together with subsequent ``Analyzing Simulated IS'') \cite{Prather2008}. Faculty may notice new aspects of these teaching strategies when simulating them as educators, and thus become better able to introspectively evaluate the fit of a particular strategy implementation to their current pedagogical beliefs and abilities, and to their local contexts.

Experiencing success at teaching during workshops could also increase faculty's self-efficacy by targeting three of the four sources of self-efficacy identified by \textcite{Bandura1986}, and thus make them more willing to experiment with new strategies in their own classrooms. If faculty perceive themselves to be successful at implementing RBIS by engaging other participants in desirable ways, they could experience a sense of mastery; if faculty perceive their peers' to be successful, they could experience vicarious success; and if faculty receive encouragement or praise for their efforts as implementors, they could be persuaded to think more highly of themselves directly (social persuasion). Faculty simulating instructional strategies themselves seems more likely to improve their confidence than watching workshop leaders do so \cite{Prather2008}: mastery experience is the most influential factor that can increase self-efficacy and cannot be achieved by watching others perform, and faculty may be more likely to experience vicarious success when watching other participants because they identify with their peers more closely.

\emph{Analyzing Simulated Instructional Strategy:}
As we discussed earlier, faculty's ability to analyze concrete examples of practice is a critical piece that can support continuing improvements to their instruction \cite{Aubusson2010,Horn2010,Morrell2012,VanEs2008,Sherin2008}. Faculty often give pedagogically superficial feedback to their peers after teaching observations, which might indicate that they also struggle to reflect on their own instruction in substantive ways. When workshop leaders vocalize their own reflective practice during workshops, they can identify alternative aspects of instruction and student engagement that may be fruitful for faculty to notice in their own classrooms. Similarly, guiding faculty to reflect on examples of instruction with their peers could improve their ability to notice consequential aspects of their students' engagement \cite{VanEs2002,VanEs2008,Goodwin1994,Prather2008,Darling-Hammond2009} and ultimately lead to shifts in their instruction (particularly if faculty continue to participate in similar activities following the workshop) \cite{VanEs2010,Sherin2008}. As discussed in Section \ref{sec:background}, scaffolded discussions could support faculty's engagement in conversational routines demonstrated to be productive within PD literature, e.g., focusing on specific challenges observed and connecting examples of classroom practice to general principles of teaching and learning \cite{Horn2010,Aubusson2010,VanEs2002,VanEs2008,VanEs2010}. Improving faculty participants' abilities to self-assess and engage in fruitful discussions about teaching episodes could have benefits that long outlast a single PD experience, and a workshop in which this code occurs frequently could generate this outcome.

\emph{Workshop Leader Pre-Workshop Experiences:}
An emphasis on a workshop leader's past experiences could lead to positive and negative workshop outcomes, depending on the context and the way that individual faculty participants perceive the workshop leader. When a workshop leader effectively emphasizes their personal development as instructors, they could give faculty participants opportunities to identify ways in which the workshop leader is or was similar to them. If faculty participants see workshop leaders as role models, they might envision themselves becoming successful at implementing research-based, student-centered instruction, thus leading to an increase in their self-efficacy \cite{Bandura1986}, as discussed in the ``Faculty Participant Simulating Instructional Strategy'' code. If faculty see the workshop leaders as peers, they may become more compelled to try the workshop leader's approach to teaching than if the same RBIS were presented without an associated personal narrative \cite{DancyPERC2010,Lund2015a}. At the same time, in physics, there is evidence that that faculty who want to improve their teaching sometimes perceive education researchers to be too dogmatic and prescriptive about how RBIS ``should be'' implemented, which contributes to negative perceptions of the field of education research and resistance to implementing or adapting RBIS \cite{HendersonDancy2008}. These negative perceptions could be generated or reinforced by an over-emphasis on a workshop leader's past experiences, particularly if the workshop leader portrays their instructional approach as inflexible or ``correct.'' More broadly, if the balance between a focus on the workshop leader's and faculty participants' past experiences is strongly skewed towards the workshop leader, this could indicate that the workshop leader's ideas and experiences are being privileged over participants' ideas and experiences within the session. Workshop designers could use R-PDOT data to identify places where they could help faculty to participate more centrally in the workshop by eliciting faculty's experiences instead of exclusively sharing their own.

\emph{Faculty Participant Pre-Workshop Experiences:}
A focus on faculty participant's past experiences could be an effective mechanism to encourage faculty to try RBIS, and could contribute to faculty becoming more reflective teachers. For the first point, faculty are most often convinced to try RBIS by their peers \cite{Lund2015a,DancyPERC2010} and feel pressure to conform to perceived teaching norms within their local contexts \cite{Hora2012,DancyPERC2010,Turpen2016}, and faculty who are already trying out progressive instructional methods might be sought out for advice about teaching by their peers \cite{Judson2007}. Discussion surrounding faculty's own teaching could foster a sense of community among participants, which is a critical part of sustaining educational reforms \cite{Darling-Hammond2009,Chappell2007,Rundquist2015}, and workshop leaders could encourage faculty participants who have already tried one or more RBIS in their classrooms to share their experiences as a way to normalize the potential challenges and advantages of research-based teaching among participants \cite{Horn2010}. For the second point, faculty naturally draw on a variety of prior experiences when making teaching decisions \cite{Oleson2013}, and it could be useful to scaffold those seeds for productive reflection by drawing them out within the workshop. For instance, if participants are encouraged to relate their past experiences to simulated classroom experiences or task design within the workshop, this could contribute to making these simulated workshop experiences seem concrete and realistic. If workshop leaders facilitate these conversations using dialogic participant structures where faculty participants voice their own ideas without the workshop leader's direct evaluation or judgment, i.e., if workshop leaders give faculty agency to recall and interpret their own experiences, these discussions could model productive conversational norms that faculty could engage in elsewhere \cite{HornLittle2010}.

%
%
%

\emph{Student Experiences:}
A focus on student experiences could lead to shifts in how faculty participants perceive their own students, which again could be a positive or negative workshop outcome depending on the nature of these shifts. Faculty's perceptions of student attitudes towards active learning and academic preparation often influence their decisions to use RBIS \cite{Lund2015a,Turpen2016}; for instance, perceived resistance to active learning is a common barrier to physics faculty implementing Peer Instruction, as is the perception that students will not have sufficient knowledge and skills to benefit from talking with their peers \cite{Turpen2016}. Because of these pervasive challenges, conversation surrounding student experiences could counteract or reinforce this deficit model, and potentially shift faculty's perceptions accordingly. Part of the purpose of including this code is to alert workshop leaders to when these conversations are occurring so that they can reflect on the prevailing attitudes towards students voiced within the workshop and work to direct these conversations towards understanding student perspectives. Productive conversations could help faculty think about how to teach a diverse student population in more equitable and inclusive ways, potentially through identifying how students' experiences may be racialized, gendered, or otherwise shaped by societal influences and considering how to constructively address these challenges \cite{Seymour2000,Halley2011,Carlone2004,Steele1997,Johnson2007}. If student experiences are considered simultaneously to workshop activities coded as ``Analyzing Simulated Instructional Strategies,'' ``Analyzing and/or Creating Student Tasks,'' and/or ``Disciplinary Content Knowledge,'' faculty participants may improve their abilities to identify, notice, and respond to students' disciplinary ideas \cite{Darling-Hammond2009}, which could in turn help them to assign appropriate tasks and act as effective facilitators in the classroom \cite{Robertson2015,Coffey2011}.

\emph{Disciplinary Content Knowledge:}
Workshops that incorporate content knowledge from participants' primary disciplines may be more salient to faculty than workshops that only discuss instructional strategies and learning theory generically \cite{Loucks-Horsley2009}. Although we see disciplinary content knowledge as involving both core ideas and cross-cutting practices \cite{NRCFramework2012}, practices often span many disciplines, and we choose to only select this code when illustrated with topics that are familiar to participants (i.e., disciplinary core ideas grounded in the disciplinary domain). Prior studies have shown that faculty's primary discipline influences how they teach and think about teaching \cite{Lund2015a,Singer1996}, which suggests that a focus on disciplinary content knowledge could help faculty to perceive the teaching strategies presented at workshops as directly relevant and more easily applicable to their own teaching. When the workshop content aligns with faculty's instructional goals, they may choose to take student tasks from the workshop and use them as building blocks in their instruction. Faculty may be better able to generate new, high-quality classroom activities when the tasks they interact with during workshops target similar knowledge or skills to the content they will teach. Also, as mentioned in the previous code, faculty's abilities to identify what knowledge and skills students possess currently and could gain through instruction can importantly shape faculty's instruction \cite{Coffey2011,Darling-Hammond2009,Robertson2015}, and activities surrounding this (e.g., examining student work or classroom video) are most likely to be valuable when they are focused on the ideas of learners within the participants' own disciplines \cite{VanEs2002}.

\emph{Analyzing and/or Creating Student Tasks:}
Analyzing and creating student tasks within workshops could improve faculty participants' abilities to engage their students with pedagogically valuable and appropriate materials in the future. Faculty constantly go through a process of selecting, modifying, and/or creating tasks when they are planning to teach, and use a variety of criteria to determine what tasks are best \cite{Hora2012,Markauskaite2014,Stroupe2013,Cohen2003}. The instructional materials faculty choose influence their instruction more broadly, which underscores the importance of this decision-making process \cite{Cohen2003,Hora2012}. Structured workshop activities that focus on student task creation and analysis could expand faculty's vision of what criteria to consider when deciding whether and in what capacity to build from existing tasks (research-based or otherwise), and what features of classroom tasks might constrain or support student learning and engagement. Student tasks could also function as the contextual anchors that ground conversations about teaching and learning in concrete, relevant examples and thus make those conversations more focused and productive \cite{Aubusson2010,Horn2010,Morrell2012}. If workshop leaders scaffold faculty's productive engagement in creating and evaluating tasks within a workshop, faculty who participate in local or virtual learning communities may become better able to initiate or contribute to productive conversations within those groups later on, in addition to whatever individual gains in ability they might have made from the workshop experience alone.  

\emph{Planning for Faculty Participant Future Teaching:}
While workshop activities represented by the other codes could help faculty participants to envision high-quality instruction and feel a sense of community within the workshop, any changes faculty are considering are more likely to be enacted if they are given opportunities to plan out next steps and think about how these changes might play out in their local contexts. Improving instruction is necessarily a long-term process, and short-term interventions like workshops can only contribute a limited amount. If faculty come into workshops with ideas about teaching and learning that are highly different from those endorsed at the workshop, they will likely struggle to achieve more desirable student outcomes without ongoing support from other educators or PD leaders. Additionally, faculty's instructional decision-making is influenced not just by how they think their students learn, but also by a host of other factors like departmental policies and incentives (or disincentives) for trying new teaching practices \cite{Hora2012,Turpen2016}. Faculty may be more prepared to face these challenges if workshop leaders both encourage faculty to consider how potential changes could fit within their constraints and to identify supports (faculty learning communities, supportive colleagues, instructional materials, etc.) that could contribute to their future learning. 

\section{\label{sec:methods}Methods}
We now turn to the practical aspects of the R-PDOT and discuss its development, validation, and use. This section is primarily aimed at other observation tool developers and future R-PDOT users; readers who are more interested in the overall outcomes of our study may wish to skip to Section \ref{sec:analysis}. The concept of a workshop observation tool emerged from private communications with Dr. Edward Prather, and as did our connections to the research settings that supported the tool's development. We attended many iterations of two of the largest disciplinary workshops for faculty, the Physics and Astronomy New Faculty Workshop \cite{Henderson2008} (6 iterations) and the Center for Astronomy Education Tier I Teaching Excellence Workshop \cite{Prather2008} (3 iterations), as we developed the R-PDOT codes. We used published descriptions of K-12 and faculty PD \cite{Garet2001a,Loucks-Horsley2009,Darling-Hammond2009,STEMWorkshopsReport2012} to ensure that the codes span the activities most likely to occur in any workshop aimed at helping faculty to improve their teaching. 

The functional form of the R-PDOT is an online interface, hosted on the Tools For Evidence-Based Action community's General Observation and Reflection Platform (GORP), which allows an observer to code by selecting and deselecting buttons continuously throughout a ``live'' workshop session\footnote{The R-PDOT interface can be accessed at \href{http://gorp.ucdavis.edu}{http://gorp.ucdavis.edu}.}. Consistent with the nature of the codes, an observer can select one type-of-engagement code and one or more focus-of-engagement code(s) at any given time. Observers can also record timestamped comments that complement or explain these coding decisions. We note that in order to select a particular code, an observer may need to assume that there is more shared purpose in the room than may actually be the case. For instance, during many types of engagement (most notably lecture and closed discussions), the only publicly visible evidence of faculty participants' thinking is discourse between a limited number of speaking actors. In the case of no counter-evidence, we default to maintaining the workshop leader's framing. We base counter-evidence for the focus-of-engagement on a small subset of participants in the observer's vicinity, and shift to other codes or add codes as appropriate. Along similar lines, faculty's type-of-engagement may conflict with the workshop leader's instructions, e.g., faculty may spontaneously begin to discuss a prompt with their peers when asked to work independently, in which case we scan the room and select the code that matches the most common participant behavior. Otherwise, we maintain the type-of-engagement established by the workshop leader.

 We find several of \textcite{Scott2006}'s guidelines for labeling communicative approaches useful in further helping observers to identify code shifts. In particular, we agree that taking an overview \cite{Bakhtin1986} is an appropriate orientation to determining what is happening in a workshop session, as all turns of talk are linked to other turns of talk and not isolated from what was occurring previously. Taking an overview can include both considering the overall teaching purpose when deciding whether or not a code is occurring, and using contextual, non-verbal cues to help establish whether or not an interaction is evaluative. This approach is particularly relevant in determining the duration of a particular focus-of-engagement and deciding the type-of-engagement during large group interactions, where potential ambiguities can be resolved by considering how the session has been unfolding. We can often make inferences about the type-of-engagement in a large group interaction based on an initial question statement, e.g., a workshop leader may pose an obviously open or closed question to the group, and, as above, we assume this type-of-engagement is maintained unless there is evidence otherwise. 

By default, our coding methods ultimately diverge from \textcite{Scott2006}'s recommendations because the R-PDOT is intended to enable users to capture workshop events in real-time (either through in situ observations or watching video played without pause), not to provide a framework for slow, iterative coding of video episodes. We agree with \textcite{Scott2006} that it is easier to identify the nature of various interactions through replaying video multiple times, and sometimes do this ourselves when qualitatively analyzing a workshop session in depth. Some of the boundaries we have created are easier to capture in a first pass than the boundaries in \textcite{Scott2006}'s framework; for example, ``Workshop Leader Lecture'' is a single code, even though, as we noted earlier, lecture could be authoritative or dialogic in principle. More fundamentally, however, the value of enabling users to capture these aspects of workshop sessions ``live'' and without multiple iterations of coding so that workshop leaders and evaluators can immediately reflect on workshop sessions, greatly outweighs the benefits of doing so with the greatest possible accuracy. Although we have strived to make the distinctions between codes as transparent as possible to R-PDOT users, we acknowledge this underlying limitation as we explore the resulting inter-rater reliability below. 

\subsection{\label{sec:IRR}Establishing inter-rater reliability}
Establishing inter-rater reliability (IRR) requires both selecting metrics to assess reliability relative to common standards set by other observation tools, and finalizing a set of codes for which to measure reliability between observers. In the following section, we describe how two IRR metrics---the Jaccard similarity score and Cohen's $\kappa$---are appropriate for and can be applied to the R-PDOT coding scheme; the ways in which we ensured that the current code names and descriptions are accessible to potential users; and the results of IRR testing between the first and second authors. We also consider the circumstances under which future users may (or may not) wish to establish strong IRR with the R-PDOT for themselves.
 
\subsubsection{The Jaccard similarity score}
The first metric we use to calculate IRR, the Jaccard similarity score, is simply the fractional observed agreement between two observers who have coded the same set(s) of data. For example, if observer A and observer B code the same workshop session and agree 85\% of the time about whether or not lecture is occurring, the Jaccard similarity score for ``WL Lecture'' would be 0.85. We calculate the Jaccard similarity score for each code individually to allow comparisons to the reliability of the Classroom Observation Protocol for Undergraduate STEM (COPUS) instrument \cite{Smith2013}, which has a similar functional form and has undergone extensive reliability testing. The standard formula for the Jaccard similarity score, applied in this context, is provided in \textcite{Smith2013}, and we follow their strategy for IRR testing wherever possible throughout this section.

One notable difference between \textcite{Smith2013}'s approach and ours is the exact method of data collection. With the R-PDOT, observers collected data continuously throughout an observation---keeping relevant codes selected until a change in the session occurred, while \textcite{Smith2013} collected data in 2-minute intervals---checking off all codes that occurred in a given time interval. Our method of data collection provides a more accurate visual and quantitative description of workshop sessions; however, in order to calculate IRR, it is necessary to bin the data. We initially chose to bin into 1-second intervals for simplicity and to ensure sufficient sampling fidelity. In doing so, we recognize that it is unrealistic to expect observers to simultaneously change codes at that level of accuracy, i.e., within the same second, particularly when they may be selecting two or more appropriate codes based on the same change in the workshop session. We therefore decided to incorporate some additional leniency into our IRR calculation, so that minor differences in the timing of selecting codes do not reduce the overall IRR scores. 

In order to determine what amount of leniency is appropriate for this coding method, we used an excerpt of workshop video to quantify how many seconds of delay between observers can typically be attributed to realistic lags in observers noticing and selecting new codes instead of real disagreement about which codes describe a particular workshop event. To calibrate this, the first author coded the same 7-minute video segment two times consecutively before any other IRR scores had been calculated, and the two sets of data were compared using the Jaccard similarity score as described above (as if the data were collected by two observers). The video segment was selected to be straightforward to code but containing many transitions between codes, such that any differences between the two iterations could be attributed to natural delays in selecting codes rather than the observer's uncertainty about code definitions. Consistent with this assumption, the two iterations of coding were quite similar initially: all Jaccard similarity scores were 0.97 or higher. For each code, we then omit any data within $x$ seconds of when that code was selected or deselected from the inter-rater reliability calculation, increasing $x$ until all codes had similarity scores of 1.0, i.e., perfect agreement. In this way, we found that omitting data that fell 2 seconds before and 2 seconds after the selection or deselection of a particular code was sufficient to bring the two observations into perfect agreement. We therefore exclude data that is 2 seconds or less away from an observed code shift in all subsequent IRR calculations.

\subsubsection{Cohen's kappa}
While the Jaccard similarity score has the benefit of being easier to interpret than Cohen's $\kappa$ and always possible to calculate, it is not predictive of how well two observers will agree in the future and thus is not a true measurement of IRR. Cohen's $\kappa$, on the other hand, takes into account the fact that two observers might agree by random chance in any given observation, and therefore yields a conservative, statistically driven estimate of how likely it is that they will agree in the future. 
 
The formula for Cohen's $\kappa$ is:
$$\kappa = \frac{p_o-p_e}{1-p_e}$$
where $\kappa$ is the probability that the two observers will agree in the future; $p_o$ is the observed proportionate agreement between the two observers; and $p_e$ is the probability of random agreement between the two observers, based on how often codes were selected by each observer.

We calculate Cohen's $\kappa$ for each focus-of-engagement code separately. This is justified because observers can select several focus-of-engagement codes at once, which implies observers must decide whether a particular code should be selected largely independently of whether or not other codes are selected. This calculation of Cohen's $\kappa$ draws on our previous calculation directly---$p_o$ is equivalent to the Jaccard similarity score---and mirrors the calculations described in \textcite{Smith2013}. In particular, like \textcite{Smith2013}, we consider agreement for a given focus-of-engagement code to indicate times when both observers have selected the same code, and times when both observers have chosen not to select that code.

The type-of-engagement coding differs from the focus-of-engagement coding in that observers are choosing between a set of mutually exclusive codes, i.e., there will always be exactly one type-of-engagement code selected at a given time. Because of this, we calculate a single $\kappa$ for all of the type-of-engagement codes together. In this case, it is unnecessary to include times when two observers agree that a particular code is \emph{not} selected when calculating Cohen's $\kappa$; in other words, only times when both observers agree that a particular type-of-engagement code is selected count in the calculation. Here,  $p_o$ is the total number of instances when the two observers select the same type-of-engagement, and $p_e$, the probability of random agreement, is given by the expression:
\begin{align*}
\sum\limits_{i} \frac{\text{\# of times code }i\text{ was selected by observer A}}{\text{\# of times A selected any type-of-engagement code}} \nonumber \\
\times \frac{\text{\# of times code }i\text{ was selected by observer B}}{\text{\# of times B selected any type-of-engagement code}} \nonumber \\
\end{align*}
We note that because of its probabilistic nature, Cohen's $\kappa$ is only useful when there is some variation in the coding of at least one observer. If there is no variation in the coding considered for a given calculation, then $\kappa$ is undefined. Similarly, if there is only variation in one observer's coding, e.g., if one observer selects a single code for an entire session, or if one observer never selects a particular focus-of-engagement code, then $\kappa$ is automatically zero for the relevant calculation. This is a logical result that indicates a lack of statistical information needed to make a reliable predictive model: all of the observed agreement is taken to be random agreement. More generally, when there is little variation in observers' coding, either because a code was selected highly frequently or highly infrequently by one or both observers, the formal probability that observers will agree by random chance is quite high. Therefore we expect $\kappa$ values to be low in those cases even when Jaccard similarity scores indicate a close agreement.

\subsubsection{Refining codes}
As we were defining the R-PDOT codes, we periodically sought and incorporated feedback from potential users in order to ensure that the version of the R-PDOT coding scheme, for which we calculate IRR, is accessible and comprehensible to users. Dr. Stephanie Chasteen, the NSF external evaluator for this project and an experienced PD presenter, pilot-tested our preliminary focus-of-engagement codes ``live'' during a workshop by informally noting examples of each code and identifying points of confusion. We made several significant changes as a result of this feedback, including adding new codes, rewriting code names, and modifying code meanings. A tenured Astronomy faculty member, PD participant, and recent adopter of RBIS,  Dr. Derek Richardson, also provided feedback on comprehension and code names when the codes were closer to their final form. In response, we modified language and code names to make them more accessible to users with a wide variety of expertise in education research.

To further improve and validate the accessibility of the R-PDOT, we conducted one-on-one interviews with workshop leaders surrounding R-PDOT data. We interviewed six presenters from the Physics and Astronomy New Faculty Workshop (NFW), incrementally introducing them to code descriptions, anonymized example data from NFW sessions, and data from their own sessions when possible. We prompted them to talk through their interpretations, and offered our own ideas when necessary to create a productive and collaborative interview setting. These interviews confirmed the general accessibility of the code names and descriptions: workshop leaders were typically able to interpret the codes meanings with little or no guidance from the interviewer. We made minor changes to some code descriptions based on one interview, which reduced the number of words in the description without changing the meanings of the codes. More substantive points of confusion raised during these interviews could be better addressed through modifications to the visual output of the tool, as discussed further below. In addition to contributing to the development and validation of the R-PDOT, these interviews provide preliminary evidence that the R-PDOT can indeed support workshop leader reflection as intended, and we describe these results briefly in Section \ref{sec:discussion}. 

\subsubsection{IRR results}
Over the course of this study, we video-recorded 64, 45-minute to 1-hour long sessions across three iterations of the NFW. Once we had established a final set of code names and descriptions, we (re)coded multiple NFW video segments to demonstrate formal IRR for the R-PDOT coding scheme. (We had coded and discussed both ``live'' and video data periodically throughout this project in order to refine this set of codes and establish joint understanding of code meanings.) The first and second author independently coded 7, 5--30 minute segments of video based on the final R-PDOT codebook, which totaled 98 minutes of data (Table \ref{tab:IRR}). The first author used her preliminary coding to select video that spans a range of workshop activities and participant structures. For each segment, she indicated a key phrase that indicated the start of the segment, the initial codes to select (since selecting a button on the R-PDOT interface marks the start of data collection), and the duration of the episode for coding. To limit biases in coding, the first and second authors coded all of these video segments independently before comparing to each other, and only the first author reviewed the preliminary coding data to choose the video segments. 

When selecting video data for this exercise, the goal was to include video segments such that each focus-of-engagement and type-of-engagement code would be present in the final, combined IRR dataset, and such that each focus-of-engagement and type-of-engagement would be similarly prevalent to allow for more direct comparison of reliability measures across codes. Despite these ideals, however, practical considerations of how the type and focus of faculty's engagement played out during the NFW created natural limitations to how evenly we were able to sample different codes. In particular, because ``WL Lecture'' was highly prevalent across the NFW, it was difficult to avoid oversampling this type-of-engagement relative to other types while still capturing a range of focus-of-engagement codes, as evident in Table \ref{tab:IRR}. Similarly, ``Workshop Instructions'' and ``Planning for FP Future Teaching'' were quite rare, and thus it was not possible to sample these codes to the same extent as other focus-of-engagement codes. (The limitation in our ability to capture extended enactment of the ``Workshop Instructions'' code will likely persist in future studies due to the nature of that code. We discuss the lack of planning for future teaching and the pervasiveness of lecture in our analysis.) 

We find that our Jaccard similarity scores for every code are all 85\%, or above which we consider to be sufficient for this study. We note that the standard set by the COPUS team \cite{Smith2013} is for raters to achieve Jaccard similarity scores above 90\%. The $\kappa$ values we include represent a lower limit on the likelihood that these two observers will agree in the future, and provides another metric for future R-PDOT observers to compare to. \textcite{Landis1977} provide some general guidance for how to interpret these results: they characterize $\kappa$ values as poor ($<$0.00), slight (0.00--0.20), fair (0.21--0.40), moderate (0.41--0.60), substantial (0.61--0.80), and almost perfect (0.81--1.00). By these benchmarks, we find that IRR for the type-of-engagement codes is moderate (0.56). We also find that the two observers simultaneously selected focus-of-engagement codes occurring for more than 10\% of the time with fair to moderate IRR, and that only focus-of-engagement codes selected for less than 5\% of the total time showed poor or slight IRR. Because those codes occurred so rarely, we should not be surprised that the reliability in selecting them was worse. Thus, we consider these results to be sensible and sufficient for our purposes, but note that these rare codes may need more refinement than other codes in the future.

Because we do not have immediate plans to explore quantitative research questions with the R-PDOT ourselves, we choose not to iterate among ourselves for higher reliability nor to pursue IRR with additional observers at this stage. As the initial developers of the R-PDOT, we primarily aim to outline the procedures needed to perform IRR calculations for this instrument in order to streamline this for future researchers. While we have established what we consider to be a viable and useful set of descriptive codes and have demonstrated that it can be used reliably within our research team, we recognize that future researchers may wish to modify the codes as the R-PDOT is tested in other settings, which would require IRR to be revisited. To the extent that future users wish to use the R-PDOT as a reflective tool only, establishing IRR and maintaining these exact codes may not be necessary: in these cases, we consider it more important that the observer and observee agree about code meanings. The Generalized Observation and Reflection Platform (GORP) where the R-PDOT is currently hosted allows users to easily create custom versions of the R-PDOT for their own use and share these custom versions with others. Based on our experiences interacting with users of classroom observation tools like the COPUS, it is reasonable to anticipate that some custom modifications and improvements could be made over time.

\bgroup
\def\arraystretch{1.5}
\begin{table*}
\caption{\label{tab:IRR} Results of inter-rater reliability testing for the authors, measured with the Jaccard similarity score in all cases and Cohen's $\kappa$ when appropriate. These calculations include 98 minutes of data comprised of 7, 5--30 minute segments of NFW sessions. Results for codes that were selected for more than 10\% of the total time coded are in bold, and results for codes that were selected for less than 5\% of the time are italicized. As justified in Section \ref{sec:IRR}, we calculate a single value of Cohen's $\kappa$ the type-of-engagement codes, and separate values for each focus-of-engagement code.}
\begin{ruledtabular}
\begin{tabular}{l c c c c}
\emph{Code name} & \emph{Minutes coded} & \emph{Fraction of total time} & \emph{Jaccard score} & \emph{Cohen's $\kappa$}\\
 & \emph{(observer average)} & \emph{(observer average)} & & \\
\hline
WL Lecture & \textbf{61.0} & \textbf{0.62} & \textbf{0.85} & --\\
LG Closed FPQ & \textbf{13.1} & \textbf{0.13} & \textbf{0.93} & ---\\
LG Closed WLQ & 7.4 & 0.08 & 0.91 & ---\\
LG Open Discuss & \emph{2.8} & \emph{0.03} & \emph{0.96} & ---\\
SG Discuss & 6.1 & 0.06 & 0.98 & ---\\
FP Present & \emph{4.9} & \emph{0.05} & \emph{0.97} & ---\\
FP Ind Work & \emph{2.6} & \emph{0.03} & \emph{0.98} & ---\\
All type-of-engagement codes & \textbf{98.0} & \textbf{1.0} & --- & \textbf{0.56} \\
\hline
Workshop Instructions & \emph{2.8} & \emph{0.03} & \emph{0.96} & \emph{0.07} \\
Education Research Theory and Results & \textbf{21.1} & \textbf{0.22} & \textbf{0.93} & \textbf{0.77} \\
IS Description and Purpose & \textbf{31.9} & \textbf{0.33} & \textbf{0.85} & \textbf{0.64}\\
WL Simulating IS & \textbf{11.4} & \textbf{0.12} & \textbf{0.90} & \textbf{0.41} \\
FP Simulating IS & 8.6 & 0.09 & 0.97 & 0.75 \\
Analyzing Simulated IS & 8.4 & 0.09 & 0.89 & 0.22 \\
WL Pre-Workshop Experiences & \textbf{11.9} & \textbf{0.12} & \textbf{0.88} & \textbf{0.36}\\
FP Pre-Workshop Experiences & 5.9 & 0.06 & 0.95 & 0.44\\
Student Experiences & 5.9 & 0.09 & 0.89 & 0.18\\
Disciplinary Content Knowledge & \textbf{23.6} & \textbf{0.24} & \textbf{0.87} & \textbf{0.60}\\
Analyzing and Creating Student Tasks & \emph{4.1} & \emph{0.04} & \emph{0.97} & \emph{0.58}\\
Planning for FP Future Teaching & \emph{1.7} & \emph{0.02} & \emph{0.97} & \emph{-0.01}\\
\end{tabular}
\end{ruledtabular}
\end{table*}
\egroup

\subsection{\label{sec:visualization}Data Visualization}
We consider it a critical piece of the R-PDOT development to link the numerical output described above to accessible, intuitive visuals. We intend the R-PDOT output to serve as a reflective anchor for workshop leaders, as well as a mechanism for researchers to identify compelling video data for future analysis quickly and easily, and this necessitates visualization design choices that take into account the nature of desirable sense-making around R-PDOT data. In this section, we describe the representations of R-PDOT data we have developed, tested, and refined based on user feedback.  

The most basic representations of the R-PDOT output are summative plots showing the percentage of the total session time spent enacting each code. We take a similar approach to the designers of the COPUS tool in this respect \cite{Smith2013}; however, we agree with \textcite{Lund2015} that although pie charts are easy to interpret quickly, they are misleading when the data represented can sum to more than 100\%. Because our type-of-engagement codes represent mutually exclusive categories and our focus-of-engagement codes do not, we use pie charts to represent the prevalence of the type-of-engagement codes and bar charts to represent the prevalence of the focus-of-engagement codes. 

\begin{figure}
\includegraphics[width=\columnwidth]{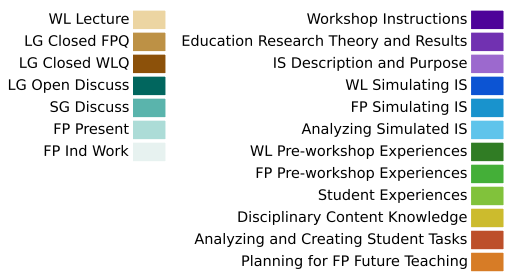}
\caption{\label{colorkey} Color key for the type-of-engagement (left) and focus-of-engagement (right) R-PDOT codes.}
\end{figure}
While we initially repeated the same color palette across the two sets of codes, in the final version, we have associated each code with a unique color, and shown in the color key (Figure \ref{colorkey}). Although this change increased the total number of colors in our representations, we found that workshop leaders consistently sought to associate codes with like colors across categories during interviews, e.g., relating a yellow focus-of-engagement code to a yellow type-of-engagement code. This mapping was unintentional on our part, and we do not wish to narrow workshop leaders' vision for how activities could be structured by implicitly encouraging particular associations. Furthermore, we found that workshop leaders were sometimes overwhelmed by the overall number of codes, and confused by which color associations were intentional and which were arbitrary. Although we have included a greater number of colors in our final version, we now only use similar colors when highlighting similarities between codes, thus simplifying the interpretive work for the user. 

The final color palette for the type-of-engagement codes, based on a ColorBrewer palette for diverging data\footnote{The original ColorBrewer (v1.0) was funded by the NSF Digital Government program during 2001--02, and was designed at the GeoVISTA Center at Penn State (National Science Foundation Grant No. 9983451, 9983459, 9983461). The design and rebuilding of the new version (v2.0) was donated by Axis Maps LLC, winter 2009 and updated in 2013.}, cleanly maps onto the four communicative approaches as outlined in Table \ref{tab:communicativeapproaches}. Brown in our color scheme is associated with authoritative communicative approaches while teal is associated with dialogic approaches, and darker shades of the same hue indicate an equal or greater level of interactivity compared to neighboring codes. We cluster the focus-of-engagement codes using a separate palette with five hues---purple, blue, green, yellow, and orange---and up to three different shades (and therefore codes) per hue. While the associations within the focus-of-engagement clusters are not as robust as the type-of-engagement associations in the sense that similarities in the research underpinnings are weaker, these clusters can help to initially orient the user to what kinds of activities are occurring and therefore reduce cognitive load when interpreting data. Specifically, Specifically, purple indicates codes that relate to describing or making sense of established knowledge; blue indicates codes that relate to in situ simulation of instructional strategies; green indicates codes that relate to the experiences of various stakeholders in the educational process; yellow indicates disciplinary content knowledge (a single code); and orange indicates codes that relate to primarily forward-looking activities. We consider the connections among the blue and green codes to be the strongest, and note that comparing the prevalence of the codes within these clusters adds to the interpretive power of the tool. These colors match the colors on the online R-PDOT interface, which allows coders to develop facility in interpreting these plots during the data collection process.

While the summative plots provide an overview of workshop sessions, which can be useful for comparing across multiple sessions and inferring potential session goals, they do not showcase the full range of information captured with the R-PDOT. Inspired by the design decisions made by \textcite{West2013} for the Real-Time Instructor Observation Tool (RIOT), we also created sets of timelines to represent R-PDOT data.\footnote{Currently, the Python scripts used to generate these timelines and all other representations shown can be obtained on the Digital Repository at the University of Maryland (DRUM) or by contacting the lead author. Ultimately, we anticipate that these visualizations will be generated automatically following data collection on the GORP website, and the original Python scripts will be downloadable from a GORP repository.} These timelines enable users to develop interpretations that rely on coordinating the type-of-engagement with the focus-of-engagement or understanding how a sequence of events unfolded. Before timelines were introduced during interviews, more than one workshop leader stated that whether or not lecture was broken up by other types of engagement mattered to their assessment, and this information is clearly visible in a timeline representation. Our interviews with workshop leaders also support the idea that timelines have particular affordances for helping presenters to reflect on their own sessions. We suspect that workshop leaders are better able to recall recent events and generate specific modifications to implement in future sessions when looking at timeline visualizations instead of an overview alone. For instance, in considering data from their own session (coded by the interviewer), one workshop leader stated
\begin{quoting}
\emph{``I like the timelines because I know what slides were happening, and so seeing how those translate into these categories I think is interesting. I could see myself, if I'm giving a lot of workshops, going through and saying, (pointing to codes on the focus-of-engagement timeline) okay this activity is hitting these bits, what happens if I do an activity that hits these bits? And then seeing what happens, what's the outcome. Because now I know I'm not doing this. So maybe I should try it and see what happens. And then I can map it into what I was actually doing (motioning to the type-of-engagement timeline).''}
\end{quoting}
As researchers who study PD workshops through multiple methods, we have also found these timelines useful in helping us to quickly select video segments for both detailed qualitative analysis (e.g., \citet{Olmstead2015}), as a representation akin to a ``content log'' in some respects, and IRR testing, as alluded to earlier.

Visually, the timelines maintain the same color scheme as the summative plots. Because the type-of-engagement codes again represent mutually exclusive categories while the focus-of-engagement codes can occur simultaneously to each other, we show the occurrence of type-of-engagement codes on a single, color-coded timeline and the focus-of-engagement codes on stacked timelines, with a single code in each row. We note that the type-of-engagement color palette we adapted from ColorBrewer is designed to be colorblind safe; thus, we can use all of these colors on a single row without strongly limiting access to colorblind users.

We have selected three sessions from the Physics and Astronomy New Faculty Workshop (NFW) as examples of how these visualizations appear, and to demonstrate what inferences can be made from them. Out of the 64 NFW sessions we video-recorded, we chose two sessions that are fairly typical of the NFW based on our experience and preliminary coding, and re-coded them from video using the finalized R-PDOT codes. These sessions allow us to illustrate some affordances and limitations of common faculty PD approaches, and will likely seem familiar to workshop leaders. We also selected a third session that contains design elements we highly value but find to be rare within the NFW. By including this session, we hope to illustrate the capacity of the R-PDOT to document a wide range of session types and to expand what workshop designers see as possible within short faculty PD sessions. We note that while we hypothesize about possible outcomes based on the empirical and theoretical work we have already introduced, we do not claim that any of these outcomes have occurred. Instead, we simply aim to illustrate the capacity of the tool to help workshop leaders and researchers to develop plausibility arguments for session outcomes and make to judgments about what post-workshop data would help to diagnose these outcomes. Our interpretations of these three workshop session follow in the next section.

\section{\label{sec:analysis}Session Analysis}
\begin{figure*}
\includegraphics[width=\textwidth]{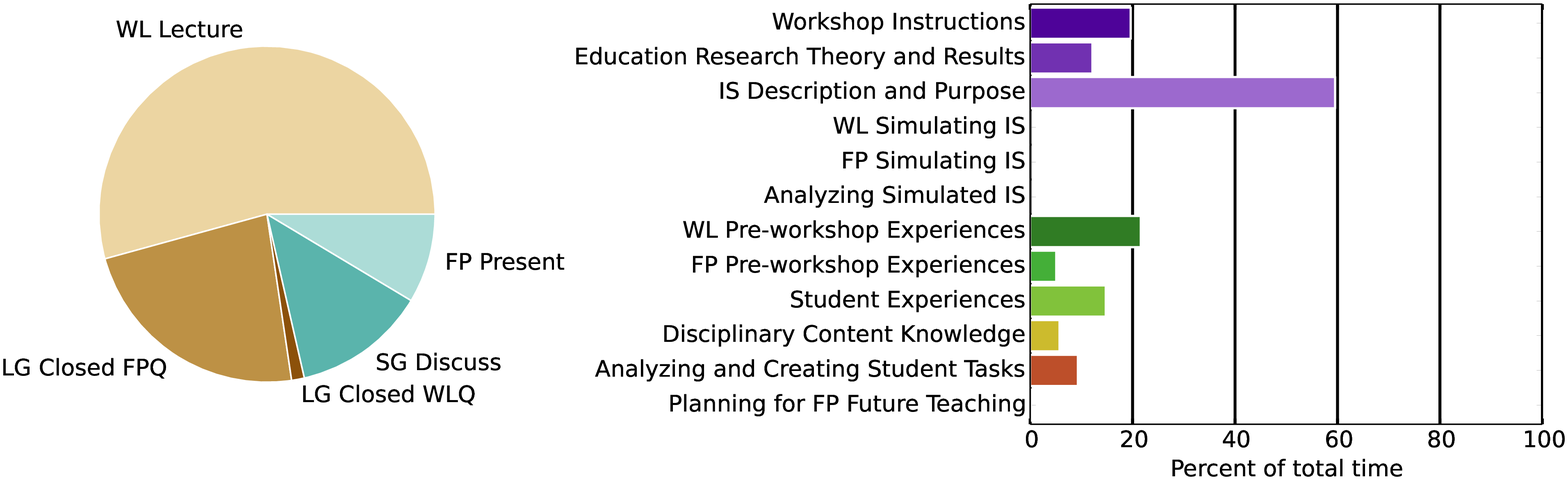}\\
\includegraphics[width=\textwidth]{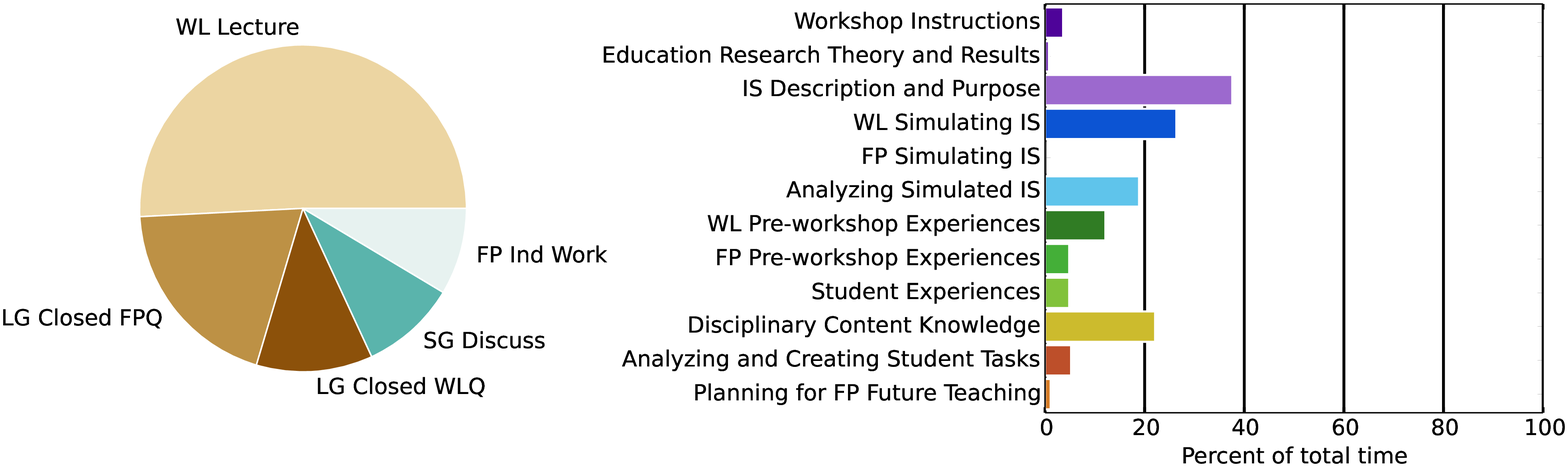}\\
\includegraphics[width=\textwidth]{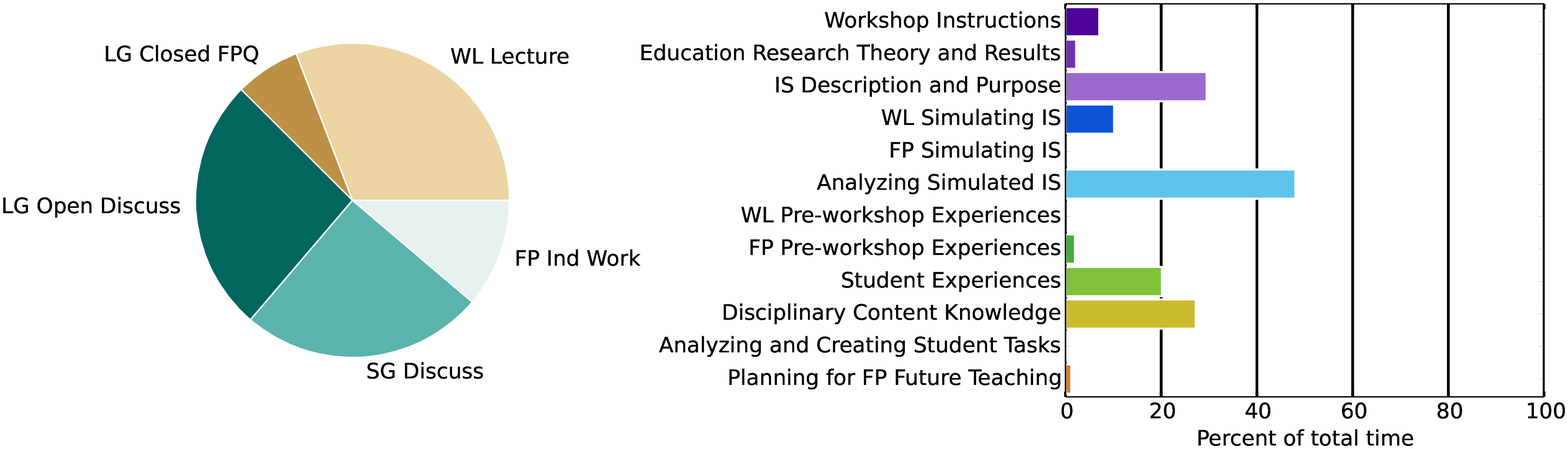}
\caption{\label{threeplots}Summative plots showing how much time was spent enacting each R-PDOT code during three approximately 1-hour long sessions of the Physics and Astronomy New Faculty Workshop, labeled A, B, and C. The pie charts show the type of faculty's engagement as defined in Table \ref{tab:typeofengagement}, where authoritative communicative approaches are shown in brown and dialogic approaches are shown in teal; the bar charts shows the focus of faculty's engagement as defined in Table \ref{tab:focusofengagement}. Further discussion of the color scheme can be found in Section \ref{sec:visualization}.}
\end{figure*}

The R-PDOT data for three sessions of the NFW---labeled Sessions A, B, and C---are shown in Figure \ref{threeplots}. Sessions similar to A and B occur regularly at the NFW, while C is less typical. Each session lasted for approximately one hour (A: 60 minutes; B: 65 minutes; C: 48 minutes). We provide our interpretations of each session based on these plots in the text that follows.

\subsection{\label{sec:sessionA}Session A}
\emph{Type-of-engagement:} Workshop leader A used authoritative communicative approaches (shown in brown) for more than 75\% of the session time. Pre-planned lecture accounted for most of this, followed by the workshop leader responding to faculty participant questions. For the $\sim20\%$ of the session that was dialogic (shown in teal), most of it took the form of small group discussion, which allowed participants to share and develop their ideas with a few other participants, and having faculty present ideas to the whole group (potentially following this small group discussion). 

\emph{Focus-of-engagement:} A majority of the session was spent focusing on previously established knowledge, particularly the description and purpose of instructional strategies ($\sim60\%$ of the time). A small amount of this discussion of established knowledge focused on education research theory and results, likely to justify or support the use of these instructional strategies. A minimal percentage of the time was spent engaging in analysis of student experiences ($\sim15\%$) and student tasks ($\sim10\%$), less than the percentage of time spent considering workshop instructions ($\sim20\%$) that may have framed those activities. The workshop leader also shared his/her personal experiences during the session, but did little to elicit or discuss participants' prior experiences. 

\emph{Potential outcomes:} Faculty would likely leave this session more aware of the RBIS discussed, and potentially motivated to use the strategies if they seemed to address pressing concerns or needs. It seems unlikely that faculty participants would gain more than a surface-level understanding of these strategies, however, and would likely not take up new ideas that conflict with their current views about teaching and learning. We claim this in part because the session was so lecture-heavy, which gave faculty highly limited opportunities to voice, reason through, or refine their current ideas. Moreover, regardless of the type of engagement, the whole session was primarily oriented towards abstract descriptions of strategies, with highly limited attempts to connect these strategies to concrete experiences within or outside of the workshop.

\subsection{\label{sessionB}Session B}
\begin{figure*}
\includegraphics[width=\textwidth]{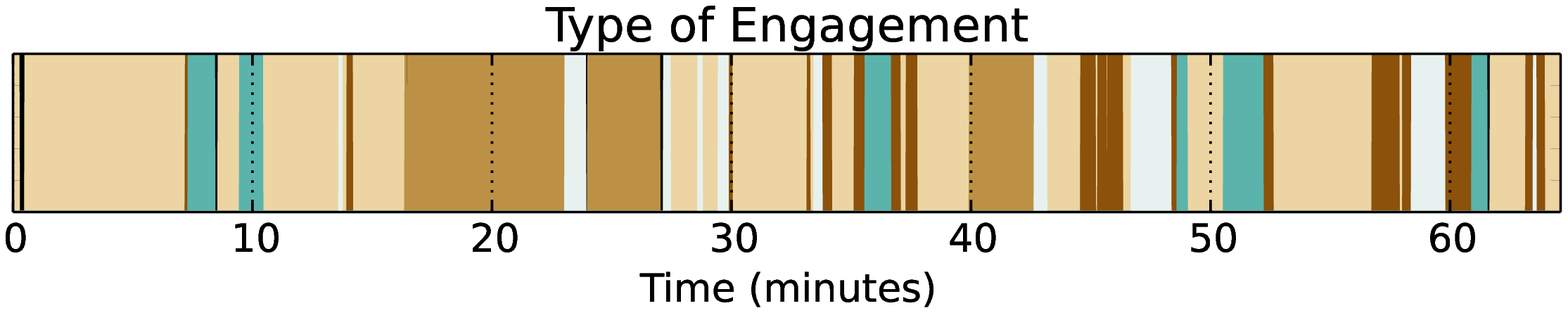}
\includegraphics[width=\textwidth]{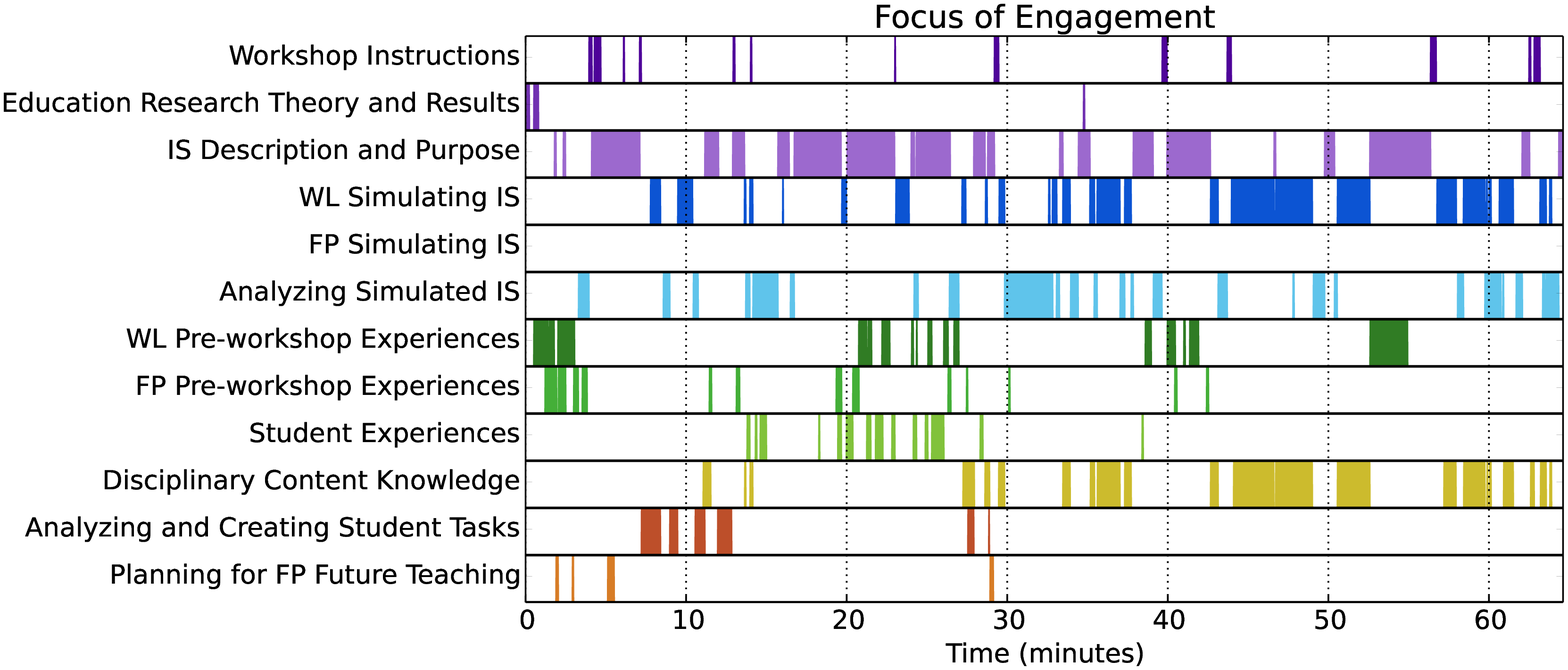}
\caption{\label{sessionBtimelines} Timeline representations of the R-PDOT data for Session B, showing how the type (top) and focus (bottom) of faculty's engagement shifted during the session. The colors are consistent with those in Figures \ref{colorkey} and \ref{threeplots}, where again, brown indicates authoritative communicative approaches and teal indicates dialogic communicative approaches for the type of engagement, and the focus-of-engagement colors indicate loose clusters.}
\end{figure*}

\emph{Type-of-engagement:} The workshop leader in Session B used authoritative communicative approaches (brown) for over 75\% of the time, as in Session A, mostly by lecturing and then by answering faculty participants' questions. Faculty were given time to think through ideas independently and could have explored divergent ideas in small groups, while all large group discussion was authoritative or closed. 

\emph{Focus-of-engagement:} Although Session B still primarily focuses on the description and purpose of instructional strategies, with some justification through research studies, the focus-of-engagement profile differs significantly from what we saw in Session A. Here, the workshop leader draws participants' focus to the concrete implementation of instructional strategies, and gives faculty the opportunity to experience these strategies as pseudo-students. There is less emphasis on the workshop leaders' past experiences, and some (albeit limited) mention of faculty and students' experiences.

\emph{Coordinating codes:} Because the focus of faculty's engagement is highly variable (especially in comparison to Session A) and seems to target more than declarative knowledge, we can learn more by considering a timeline representation showing how the events in Session B unfolded (Figure \ref{sessionBtimelines}). In carefully examining these timelines, we make two key observations about the nature of the interactions in the session: first, that all of the dialogic types of engagement occurred when faculty were experiencing well-defined RBIS, typically in conjunction with disciplinary ideas, and second, that all of the analysis of the in situ teaching events was dictated by the workshop leader, either through lecture or short periods of closed questioning. We also find that although the description and purpose of instructional strategies is the most prevalent code, the workshop leader simulates instructional strategies and analyzes their implementation intermittently throughout the entire session, indicating to us that this was a consistent theme and likely salient to participants.

\emph{Potential outcomes:} Modeling instruction for faculty participants may help them to envision changes to their practice that would be difficult to identify in the abstract, and therefore this session could be more successful than Session A at shifting faculty's instruction towards a student-centered model. The workshop leader demonstrated for faculty how they could justify the instructional choices they experienced as students, which might improve faculty's ability to reason about these specific strategies. We anticipate that faculty would likely make well-reasoned choices about whether to adopt or reject the workshop leaders' suggestions based on their incoming pedagogical understanding and how well their goals align with the workshop leader's justifications. Some faculty participants might also use the workshop leader's reflection on their practice as a model for how to engage in reflective practices outside the classroom more generally. But because the workshop leader presented these analyses of instruction through authoritative communicative approaches, there were limited opportunities for faculty to develop their abilities along this dimension within the session.

\subsection{\label{sessionC}Session C}
\begin{figure*}
\includegraphics[width=\textwidth]{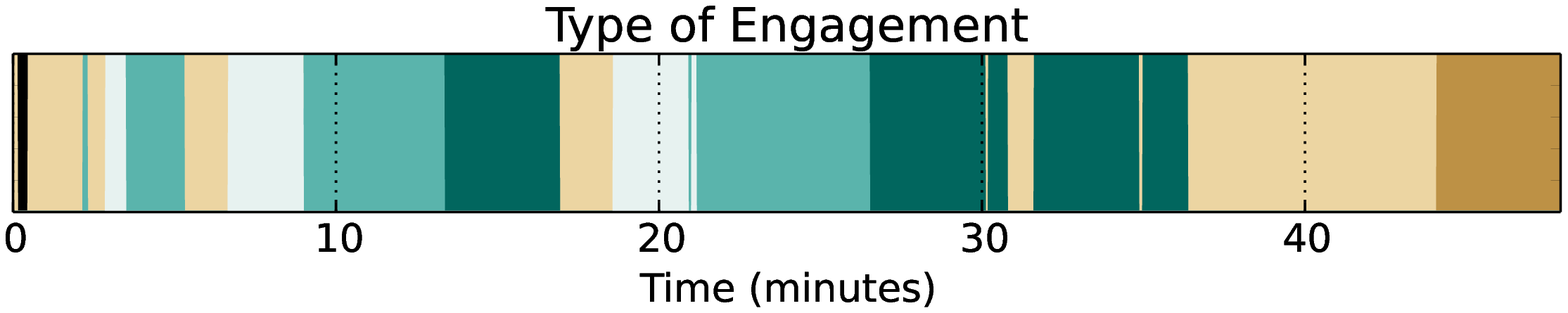}
\includegraphics[width=\textwidth]{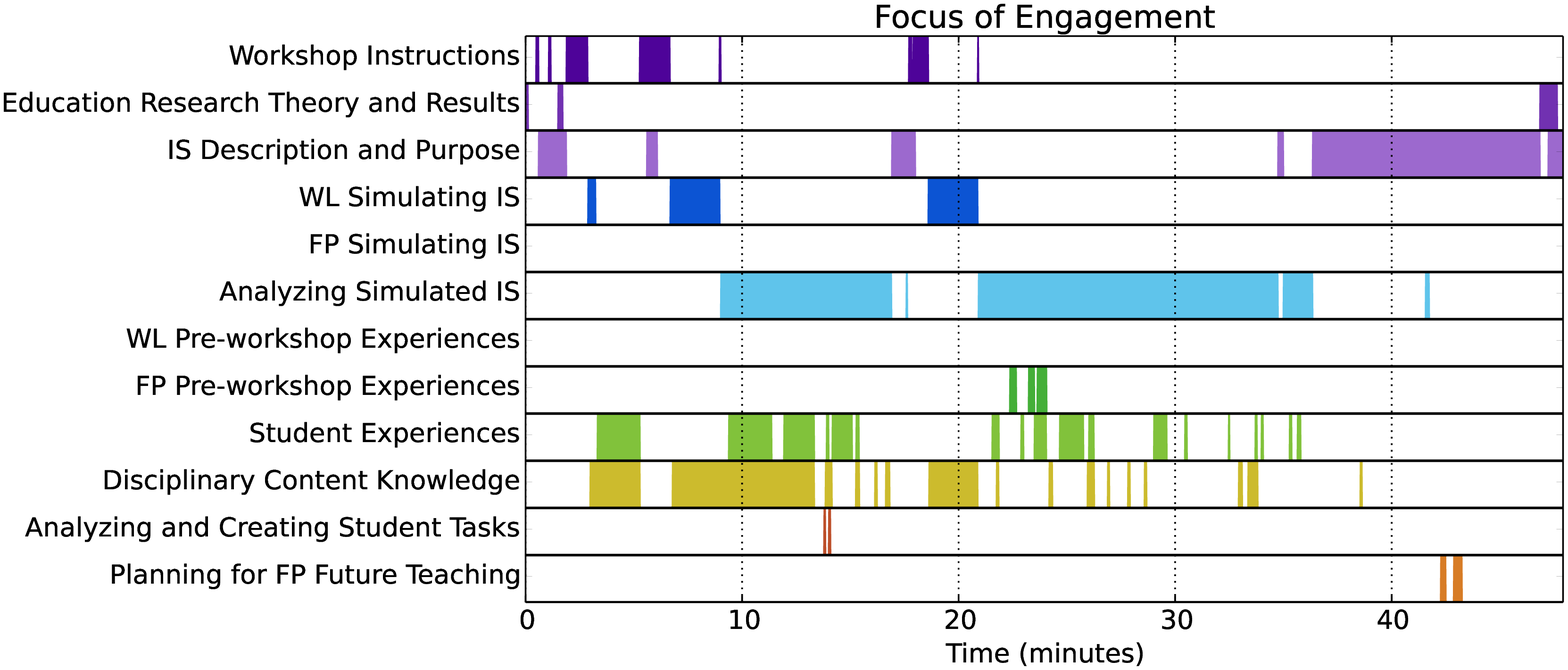}
\caption{\label{sessionCtimelines} Timeline representations of the R-PDOT data for Session C.}
\end{figure*}

\emph{Type-of-engagement:} The type of engagement in Session C is mostly dialogic ($\sim60\%$), including a quarter of the session time spent with participants engaged in both large group open discussion and small group discussion. There are certainly some authoritative segments in Session C, primarily lecture and some responding to faculty questions, but the structure of the session overall dramatically contrasts with the authoritative approaches of workshop leaders A and B. This distribution of types of engagement seems to indicate that workshop leader C is primarily focused on helping faculty participants to develop their own ideas as opposed to prescribing a single approach.

\emph{Focus-of-engagement:} The primary activity across the entire session is analyzing simulated instructional strategies. Student experiences are included in this discussion, whereas the workshop leaders' experiences are never considered and faculty's experiences are considered very rarely. As in Session B, the time that workshop leader spends focusing on the description and purpose of instructional strategies is greatly reduced relative to Session A, and instead, the workshop leader focuses faculty participants on concrete examples of instructional strategies simulated in situ. 

\emph{Coordinating codes:} Looking at the timelines for Session C (Figure \ref{sessionCtimelines}) provides information about how these activities are structured and coordinated. We can see that the workshop leaders' simulation of instructional strategies happens in two discrete chunks with faculty participants working individually, which suggests that faculty may be watching a scenario being enacted as opposed to participating in it as pseudo-students. In contrast to Session B, faculty are made responsible for the complex task of analyzing the simulated instructional strategies: all of the reflection and analysis occurs within interactive, dialogic participant structures. We also note that the final 10--15 minutes of Session C take on a different structure and focus than the rest of the session, namely, the workshop leader describing instructional strategies through lecture and answering faculty questions. 

\emph{Potential outcomes:} The outcomes we might expect from Session C are quite different from Sessions A and B. Because we see faculty first being given agency to analyze instruction with a small subset of their peers, and then further developing and comparing the many ideas that may have emerged in these small group discussions with the workshop leader's subtle guidance, we find it highly plausible that this session could improve faculty's ability to reason about new pedagogical situations, notice new aspects of their students' thinking, and consider a diversity of potential teaching choices. Faculty might also become better able to initiate and engage in these kinds of pedagogical discussions with other educators in the future. On the other hand, Session C does not link directly to faculty's prior teaching experiences or project into their future teaching. These factors combined with the lack of a prescribed instructional model (as in Session B) may make it difficult for faculty to identify concrete next steps for improving their instruction based on this session. The dialogic nature of this session may also challenge faculty's expectations about the workshop session, e.g., faculty who expect that they will be told how to teach may perceive the session to be less valuable, and result in lower evaluations compared to more prescriptive sessions.\footnote{On a post-workshop evaluation survey, faculty's ratings of the ``usefulness'' of Session C was lower than for many other sessions at that iteration of the NFW.}

\section{\label{sec:discussion} Discussion}
The information captured with the R-PDOT can be used to consider the merits and limitations of sessions taken individually, as demonstrated in the previous section, and to consider workshop design more holistically. For the two sessions we present as representative of many others within the Physics and Astronomy New Faculty Workshop (NFW) (Sessions A and B), we find that dialogic types of engagement are highly limited, and that presenters rely heavily on lecture. In contrast, dialogic types of engagement are highly prevalent in the one session (Session C) that is anomalous within the NFW. These three sessions also illustrate that although opportunities for faculty to consider the description and purpose of instructional strategies occur regularly in NFW sessions, opportunities for faculty to interactively experience concrete examples of practice are a central focus in some sessions (like Sessions B and C) but highly limited or non-existent in others (like Session A). We consider lecturing about instructional strategies likely to be a successful mechanism for raising faculty's awareness of what RBIS exist, which is highly consistent with prior research on NFW outcomes \cite{Henderson2012}, and we consider these increases in awareness to be beneficial to an extent. In particular, we think that faculty who already possess extensive pedagogical knowledge or participate in faculty learning communities could most easily take advantage of new resources in ways that genuinely enhance their students' learning and engagement. On the other hand, we find it difficult to imagine how faculty who are less knowledgeable or experienced with research-based teaching would productively assess and enact new instructional strategies based on sessions that do not allow them to interact with these strategies in some way. Because of this, we consider concrete, interactive examples to be a critical component of PD workshops that can enable all faculty to clearly envision changes to their instruction, and have shown how we can identify these instances with the R-PDOT.

The R-PDOT data can also reveal important aspects of whether and how these concrete examples of practice, when present, are explored and assessed within workshop sessions. Both Sessions B and C spend significant time analyzing instructional strategies simulated in situ, which likely indicates that these workshop leaders have goals beyond just raising faculty's awareness of RBIS. Of the two sessions, only Session B models reflective teaching, while only Session C extensively engages faculty in evaluating teaching decisions for themselves. We find significant value in workshop leaders modeling productive reflection because they are likely adept at weaving education research theory into their analysis and noticing key events. But we argue that it is also important for faculty to have some experiences generating and refining pedagogical ideas for themselves (i.e., as learners about teaching) if we expect them to enact these dialogic/interactive approaches within their own classrooms. Moreover, as stated earlier, we do not consider the benefits of a balance between authoritative and dialogic communicative approaches to apply exclusively to teaching science content, and we consider guiding faculty to make well-reasoned but autonomous decisions an important characteristic of workshops that could contribute to faculty's long-term success. Our data suggests that workshop leader C's approach may be rare within faculty PD workshops, and we argue that this warrants further consideration from workshop leaders in the future.

The lack of occurrence of focus-of-engagement codes can also inform workshop design. While it would likely be unproductive to attempt to enact all of the focus-of-engagement codes in a single 1-hour session, individual PD facilitators could generate potential modifications to their own sessions based on codes that never occur, and PD organizers could consider what new sessions would best complement existing ones within a longer workshop. For example, when looking across these three sessions, we see that the ``Planning for Faculty Participant Future Teaching'' and ``Faculty Participants Simulating Instructional Strategies'' codes nevWith the R-PDOT, PD leaders can gather concrete evidence to use as they discuss ways to continually improve their workshops, such as ways to incorporate more design structures that could prepare faculty to adapt instructional strategies for their unique situations and motivate faculty to participate in long-term PD efforts.er occur, and the ``Education Research Theory and Results,'' ``Faculty Pre-Workshop Experiences,'' and ``Analyzing and Creating Student Tasks'' codes occur rarely. These gaps primarily relate to activities that could help faculty to fit RBIS into their local contexts, which is known to be a limitation of faculty PD more broadly \cite{Dancy2009PERC}. Encouragingly, sessions that target these activities have been added to the NFW while we have been conducting this research, and other disciplinary faculty workshops enact them as well \cite{STEMWorkshopsReport2012}. But it is unclear to us how central these activities are or will become for participants in each of these workshops, and how different disciplinary workshops compare to each other. With the R-PDOT, PD leaders can gather concrete evidence to use as they discuss ways to continue to improve their workshops to better prepare faculty to adapt to their unique situations and participate in long-term PD efforts.

Through our interviews with workshop leaders, we have found preliminary evidence that R-PDOT data can indeed support others' engagement in the type of reflection and analysis that we have demonstrated here. For example, we find that looking at R-PDOT data encourages workshop leaders to think about and articulate potential limitations of lecture. For example, while reflecting on some example sessions from the NFW (similar to Sessions A and B), one workshop leader states: 
\begin{quoting} \emph{``I think the limitation [of lecture]--I think there's shared experiences, there's experiences that each person has in the classroom or in the group that others can learn from. And here we're supposing that the person at the front is the expert and knows that, and they very well may be, but there are some experiences that can probably add greatly to the learning of everybody who's participating. And so if there's not a lot of facilitation of that engagement of the broader group then you could miss out on those experiences. And things sometimes organically grow, you know, sometimes the conversation goes in a place that you never thought it would go and it's an even better experience for everybody.''} \end{quoting}

We have also found that workshop leaders may generate and justify potential changes to their own sessions based on codes they are not yet enacting. In reflecting on their own session after reading through the R-PDOT codes and seeing example data, one workshop leader realized how to incorporate pedagogical ideas from their physics instruction into their workshop session, stating: 
\begin{quoting} \emph{``One of the things that I would do differently that it would take the exact same amount of time is what I actually do in my classes, which is, for [a specific physics prompt], instead of asking them individually [i.e., only calling on individual participants], to have them think about it for a minute and then talk amongst themselves, because instead of hearing from one person that there were two ways of doing it, likely working together they would be less afraid--just like our students--of responding, and likely come up with two or three [ways] and be less afraid to respond. … I can tell you from looking at the taxonomy just as with when I've looked at [the] RTOP and the COPUS, that you sit there and you go `Oh if I--'  and it's like `Oh it's just a minor change to make it more interactive or different.' ''}\end{quoting}

\section{\label{sec:conclusions} Conclusions}
As a community of researchers who want to help bring about improvements in undergraduate STEM education, we need to critically evaluate our own professional development practices for faculty. Our observational tool, the Real-time Professional Development Observation Tool (R-PDOT), contributes to that goal by illuminating and allowing others to document key PD practices currently used within workshops aimed at improving faculty's instruction. The R-PDOT can give a holistic sense of how interactive and prescriptive workshops are by capturing the ways in which faculty are engaged and what they are focused on. Initial empirical results from the Physics and Astronomy New Faculty Workshop suggest that faculty PD can be lecture-heavy, non-interactive, and authoritative, leaving little space for faculty to grapple with their incoming ideas about teaching and learning or their instructional contexts. If we want workshops to contribute to the pursuit of ambitious learning goals for faculty, beyond building initial awareness and interest, further shifts in PD practices will be needed. 

We intend our discussion of PD activities in this paper to serve a broad audience. In exploring these PD activities and ways of engaging faculty in thinking about teaching and learning, we hope to expand our community's vision for what faculty PD could look like. In addition to drawing attention to the interactivity and prescriptiveness of workshop sessions, we encourage increased attention to forward-looking activities (i.e., analysis of student tasks, analysis of pedagogical approaches, and planning for future instruction), lending weight to promising features of PD that are already being enacted in some PD settings. For workshop leaders who choose to take up the tool to document their own practices, the R-PDOT output can generate critical, reflective discussions surrounding aspects of PD that are likely to be consequential for workshop outcomes, and thus meaningfully inform workshop leaders' future efforts. The R-PDOT can also fuel future research and inspire us to ask new questions about how we can best support faculty learning, such as ``What does it look like for faculty to productively engage in developing course content together?,'' ``How do faculty PD workshops vary across disciplines, or when workshop facilitators have different kinds of expertise?,'' and ``How do facilitators elicit and help faculty to build from their own experiences?'' For researchers, the R-PDOT can provide baseline documentation that would enable the full pursuit of such questions.

Methodologically, the R-PDOT could be used to support future research in a variety of ways. In quantitative research, the R-PDOT could allow correlations to be made between workshop outcomes captured by survey, interview, or classroom observation data and workshop design when used across multiple sessions or workshops, thus furthering our understanding of how these design decisions are consequential to faculty. For example, the R-PDOT data highlights unanswered questions about the importance of incorporating disciplinary content knowledge into PD, which could be explored by searching for correlations between the prevalence of this code and relevant survey responses. In qualitative research, the R-PDOT data can lead users to ask targeted questions about the details of workshop design and implementation that are related to but not captured by these broad codes, such as questions about workshop leader facilitation moves or faculty interactions. These questions could be pursued by using the R-PDOT to identify episodes for detailed analysis, and coordinating with additional data sources such as workshop video, field notes, or workshop artifacts. 

Taking a broader perspective, researchers could also continue to pursue how and to what extent R-PDOT data acts as formative feedback for workshop leaders through additional interviews, discussions, and continued documentation. Such research could strengthen efforts to use observation tools as a catalyst for teacher reflection within PD and elsewhere, and help to unpack some current orientations to faculty PD in ways that could inform ongoing conversations about how members of the discipline-based education research community can effectively contribute to increasing the impact of research-based teaching innovations.

\begin{acknowledgments}
The authors would like to thank Dr.'s Edward Prather, Derek Richardson, Stephanie Chasteen, and members of the University of Maryland Physics Education Research Group for their feedback and direct contributions to this paper. We would also like to thank Dr. Robert Hilborn and the other workshop leaders at the Physics and Astronomy New Faculty Workshop and the Center for Astronomy Education Tier I Teaching Excellence Workshop who allowed us to test out our tool in their sessions and discussed workshop design with us; Cassandra Paul, Andrew Reid, and the UC Davis Tools for Evidence-Based Action team for their support in developing an online interface for the R-PDOT; and the instructors who discussed their workshop experiences with us and helped to guide our intuitions about plausible potential outcomes. Finally, we would like to the undergraduate students and AAPT staff who helped to ensure that our data collection went smoothly. This work is supported by funding from NSF-DUE 1431681.
\end{acknowledgments}
\bibliographystyle{apsrev4-1.bst}
\bibliography{library.bib}

\vfill
\begin{widetext}
\pagebreak
\appendix{}
\section{\label{appendixA}Extended R-PDOT codebook: Type-of-engagement}

\bgroup
\def\arraystretch{1.5}
\begin{table*}[!h]
\begin{ruledtabular}
\begin{tabular}{p{0.25\textwidth}p{0.25\textwidth}p{0.45\textwidth}}
\emph{Code name} & \emph{Code description} & \emph{Example observed behaviors and actions}\\
\hline
Workshop Leader Lecture & Workshop leader lectures while faculty participants listen. & -Workshop leader lecturing on a pre-determined topic. \newline -Workshop leader asking a rhetorical question.\\
\hline
Large Group Closed,\newline Faculty Participant Question & Large group closed discussion. Faculty participant(s) question the workshop leader, and (optionally) the workshop leader answers through lecturing, while all other participants listen. & -Faculty participants challenging the workshop leaders' ideas. \newline
-A workshop leader engaging in conversation with a questioner(s), with little or no participant-participant interaction. \newline
-A workshop leader addressing a participant's question at length, i.e., the workshop leader's turns of talk are significantly longer than faculty participants' turns of talk.\\
\hline
Large Group Closed,\newline Workshop Leader Question & Large group closed discussion. Workshop leader asks non-rhetorical closed questions, and (optionally) one or more faculty participants respond directly to the workshop leader\cite{Dancy2007}. & -Faculty participants voting in response to a multiple-choice question. \newline
-Participants contributing in short phrases or full sentences in response to a workshop leader's question. \newline
-Workshop leader and a few faculty participants taking turns speaking, with the discourse focused on revealing an answer the workshop leader is seeking. \newline
-A workshop leader revoicing participant responses. \newline
-A workshop leader intentionally making an implementation ``error'' and expecting participants to pause him/her. \newline
-A workshop leader reading a question or prompt aloud. \newline
\\
\hline
Large Group Open Discuss & Large group open discussion. Workshop leader and faculty participants take turns speaking, with the discourse focused on the ideas of faculty participants\cite{Dancy2007}. & -Faculty participants speaking in complete sentences and contributing new ideas. \newline
-A workshop leader asking open-ended questions and facilitating discussion between faculty participants. \newline
-Workshop leader(s) participating in discussion in a non-authoritative manner, e.g., sharing their own experiences without privileging them over participants' experiences. \newline
\\
\hline
Small Group Discuss & Faculty participants discuss with each other in small groups. & -Faculty discussing with their peers during Peer Instruction. \newline
-Faculty collaborating on assigned tasks. \newline
-Faculty discussing ideas about teaching and learning in response to a prompt. \newline
-A workshop leader giving brief instructions while most faculty are still discussing with each other. \newline
\\
\hline
Faculty Participant Present & One or more faculty participants present to all others. & -Faculty participants reporting out after working on a task in small groups.\newline
-Faculty participants reporting out after working independently.\newline
-Workshop leader facilitating presentations, e.g., calling on the next group to speak or revoicing faculty's contributions.
\\
\hline
Faculty Participant\newline Independent Work & Faculty participants work independently on a task. & -Faculty participants silently reading a Peer Instruction question. \newline
-Faculty participants watching classroom video. \newline
-Faculty participants writing independently in response to a prompt or task. \newline\\
\end{tabular}
\end{ruledtabular}
\end{table*}
\egroup

\section{\label{appendixB}Extended R-PDOT codebook: Focus-of-engagement}
\bgroup
\def\arraystretch{1.5}
\begin{table*}[!h]
\begin{ruledtabular}
\begin{tabular}{p{0.25\textwidth}p{0.25\textwidth}p{0.45\textwidth}}
\emph{Code name} & \emph{Code description} & \emph{Example observed behaviors and actions}\\
\hline
Workshop Instructions & Workshop leader instructs participants about what they should or will be doing during the workshop (or participants attempt to clarify these instructions). & -Workshop leader describing the intended purpose of the workshop or workshop session. \newline
-Workshop leader telling faculty how to act or what to focus on during an upcoming workshop task. \newline
-Faculty recounting or attempting to clarify the instructions for a workshop task. \newline
\\
\hline
Education Research Theory and Results & Workshop leader and/or participants emphasize discipline-based education research processes, principles, or findings. & 
-Describing student misconceptions identified in research. \newline
-Showing evidence of improved student outcomes in active learning environments. \newline 
-Discussing the implications of active learning for diverse student populations. \newline
-Describing or making sense of education research methods and results. \newline
-Describing the characteristics or demographics of students within a particular study. \newline
-Describing or discussing research-motivated principles in relation to teaching decisions. \newline
\\
\hline
Instructional Strategies (IS) \newline Description and Purpose & Workshop leader and/or participants show or describe active learning strategies ranging from current faculty practices to strongly research-based instructional strategies. & -Listing and describing research-based instructional strategies. \newline
-Providing instructions about where to find existing materials or questions. \newline
-Showing the specific steps that comprise an active learning instructional strategy. \newline
-Describing or discussing the purpose of any active learning instructional strategy or components within it. \newline
-Describing or discussing the functionality associated with research-based educational technologies (such as PhET sims). \newline
-Displaying and briefly discussing research-based questions or tasks. \newline
-Describing or discussing possible outcomes of using a particular research-based instructional strategy. \newline
\\
\hline
Workshop Leader (WL) \newline Simulating IS & Faculty participants experience a workshop leader's implementation of an instructional strategy, either by acting as mock students, or through observing video, transcript, or case study narrative. & -A workshop leader assuming the role of a science instructor to implement a research-based instructional strategy. \newline
-A workshop leader assuming the role of a science instructor to demonstrate common faculty practices. \newline
-A workshop leader showing classroom video. \newline
-Faculty participants reading a detailed classroom case study. \newline
-Faculty participants working through a science task as mock students. \newline
\\
\hline
Faculty Participant (FP) \newline Simulating IS (as educator) & A predetermined subset of faculty participants (one or more) try out implementing an instructional strategy while other participants act as mock students. & -Individual faculty participants implementing a research-based instructional strategy with others acting as students. \newline 
-Teams of faculty participants implementing a research-based instructional strategy with others acting as students. \newline
-Faculty participants rehearsing the implementation of a research-based instructional strategy in small groups. \newline
\\ 
\end{tabular}
\end{ruledtabular}
\end{table*}
\egroup

\bgroup
\def\arraystretch{1.5}
\begin{table*}[!h]
\begin{ruledtabular}
\begin{tabular}{p{0.25\textwidth}p{0.25\textwidth}p{0.45\textwidth}}
Analyzing Simulated IS & Workshop leader and/or participants reflect on (analyze, critique, evaluate, justify) a shared experience of someone simulating an instructional strategy in situ. & -Analyzing how students responded to a given instructor move or reasoned through a science task. \newline
-Analyzing the behaviors and actions of an instructor observed during the workshop session. \newline
-Reflecting on the quality of faculty participants' engagement in the workshop. \newline 
-Predicting how students would respond to alternative instructor moves compared to those seen or experienced. \newline
-Generating possible next steps an instructor could take following a scenario experienced or considered in the workshop. \newline
-Analyzing a hypothetical situation that is closely related to, and generated in relation to, an in situ implementation of an instructional strategy. \newline
\\
\hline
WL Pre-Workshop Experiences & Workshop leader and/or participants discuss a workshop leader's past experiences, including instructional goals, practices, values, and local contexts. & -A workshop leader describing their involvement in discipline-based education research and prior studies they have done, using first-person narrative. \newline
-A workshop leader describing their teaching practices. \newline 
-Participants eliciting a workshop leader's past experiences, such as the details of how they teach and the characteristics of their students. \newline
\\
\hline
FP Pre-Workshop Experiences & Workshop leader and/or participants reflect on participants' past experiences, including instructional goals, practices, values, and local contexts. & -Faculty participants describing specific teaching strategies they have implemented in their classrooms. \newline
-A workshop leader eliciting or stating assumptions about faculty participants' prior teaching and learning experiences. \newline
-Faculty participants describing their local teaching resources or constraints. \newline
\\
\hline
Student Experiences & Workshop leader and/or participants consider students' knowledge, skills, or affect. & -Hypothesizing about or describing known student explanations of science ideas. \newline
-Analyzing real student discourse or student work. \newline
-Describing student resistance or buy-in to different instructional approaches and discussing possible causes. \newline
-Discussing students' disciplinary knowledge and skills, and considering implications for instruction. \newline
-Describing student demographics or characteristics \newline
\\
\hline
Disciplinary Content \newline Knowledge & Workshop leader and/or participants consider disciplinary ideas. & -Faculty participants engaging in disciplinary tasks using skills and knowledge that students would likely use. \newline
-Faculty participants making sense of a task using a broad range of skills and knowledge (potentially including advanced disciplinary ideas). \newline
-A workshop leader lecturing or asking questions that require knowledge from the participants' own discipline, e.g., as part of simulating a research-based instructional strategy. \newline
\\
\hline
Analyzing and/or Creating \newline Student Tasks & Participants create, modify, or evaluate and/or workshop leaders critique or evaluate specific materials, questions, or tasks for students. & -Writing new questions during the workshop. \newline
-Critiquing existing questions or tasks. \newline
-Highlighting effective aspects of existing specific research-based materials, questions, or tasks. \newline 
-Classifying questions using Bloom's taxonomy. \newline
-Exploring the pedagogical affordances or constraints of a specific instructional strategy simulation. \newline
\\
\hline
Planning for FP \newline Future Teaching & Workshop leader advises and/or faculty participants plan next steps for when participants go back to their home institutions. & -Discussing possible ways to adapt research-based instructional strategies to fit participants' local contexts. \newline
-Advising faculty participants about how to approach the process of changing their teaching. \newline
-Faculty participants identifying questions or tasks that are particularly relevant for their own instructional goals. \newline
\\
\end{tabular}
\end{ruledtabular}
\end{table*}
\egroup

\end{widetext}
\end{document}